\def\year{2018}\relax
\documentclass[letterpaper]{article} %DO NOT CHANGE THIS
\usepackage{aaai18}  %Required
\usepackage{times}  %Required
\usepackage{helvet}  %Required
\usepackage{courier}  %Required
\usepackage{url}  %Required
\usepackage{graphicx}  %Required
\usepackage[utf8]{inputenc} % allow utf-8 input
\usepackage[T1]{fontenc}    % use 8-bit T1 fonts
\usepackage{url}            % simple URL typesetting
\usepackage{booktabs}       % professional-quality tables
\usepackage{amsfonts}       % blackboard math symbols
\usepackage{nicefrac}       % compact symbols for 1/2, etc.
\usepackage{microtype}      % microtypography

\usepackage{rotating}
\usepackage{algorithm}
\usepackage[noend]{algorithmic}
\usepackage{verbatim}
\usepackage{paralist}
\usepackage{amsmath}
\usepackage{amsthm}
\usepackage{latexsym}
\usepackage{multirow}
\usepackage{array}
\usepackage[T1]{fontenc}
\usepackage{xspace}
\usepackage{subfig}
\usepackage{color}
\newtheorem{theorem}{Theorem}
\newtheorem{corollary}[theorem]{Corollary}
\newtheorem{lemma}[theorem]{Lemma}

\newcommand{\argmax}{\mbox{argmax}}

\newcommand{\eps}{\varepsilon}

\newcommand{\ind}{{\cal I}}
\newcommand{\cI}{{\cal I}}
\newcommand{\cM}{{\cal M}}

\newcommand{\RR}{{\mathbb R}}

\newcommand{\SLS}{\textsc{{Streaming Local Search}}\xspace}
\newcommand{\LS}{\textsc{{Local Search}}\xspace}
\newcommand{\pstream}{\textsc{{IndStream}}\xspace}
\newcommand{\pdstream}{\textsc{{IndStreamDensity}}\xspace}
\newcommand{\opt}{\textsc{{OPT}}\xspace}
\newcommand{\AlgF}{\textsc{{Fantom}}\xspace}
\newcommand{\INPUT}{\item[{\bf Input:}]}
\newcommand{\OUTPUT}{\item[{\bf Output:}]}

\newcommand{\citet}[1]{\citeauthor{#1}~\shortcite{#1}}
\newcommand{\citep}{\cite}

\def\b1{{\bf 1}}

\begin{document}
\title{Streaming Non-monotone Submodular Maximization: \\ Personalized Video Summarization on the Fly}
\author{
	Baharan Mirzasoleiman \\ %Department of Computer Sciencne \\ 
	ETH Zurich, Switzerland \\ baharanm@ethz.ch
	\And Stefanie Jegelka %\\ Electrical Engineering \& Computer Science 
	\\ MIT, United States \\ stefje@mit.edu
	\And Andreas Krause %\\ Department of Computer Sciencne 
	\\ ETH Zurich, Switzerland \\ krausea@ethz.ch
}

\maketitle
\begin{abstract}
The need for real time analysis of rapidly producing data streams (e.g., video and image streams) motivated the design of streaming algorithms 
that can efficiently extract and summarize useful information from massive data ``on the fly''. Such problems can often be reduced to maximizing a submodular set function subject to various constraints. 
While efficient streaming methods have been recently developed for monotone submodular maximization,
in a wide range of applications, such as video summarization, the underlying utility function is non-monotone, and there are often various constraints imposed on the optimization problem to consider privacy or personalization.
We develop the first efficient single pass streaming algorithm, \SLS, 
that for any streaming monotone submodular maximization algorithm with approximation guarantee $\alpha$ under a collection of independence systems $\ind$, provides a constant $1/\big(1+2/\sqrt{\alpha}+1/\alpha +2d(1+\sqrt{\alpha})\big)$ 
approximation guarantee for maximizing a non-monotone submodular function under the intersection of $\ind$ and $d$ knapsack constraints.
Our experiments show that for video summarization, our method runs more than 1700 times faster than previous work, while maintaining practically the same performance.
\end{abstract}

\section{Introduction}
Data summarization--the task of efficiently extracting a representative subset of manageable size from a large dataset--has become an important goal in machine learning and information retrieval. 
Submodular maximization has recently been explored as a natural abstraction for many data summarization tasks, including image %and video 
summarization \cite{tschiatschek2014learning}, scene summarization \cite{simon2007scene}, document and corpus summarization \cite{lin2011class}, active set selection in non-parametric learning \cite{mirzasoleiman2016distributed} and training data compression \cite{wei15}. 
Submodularity is an intuitive notion of diminishing returns,
stating that selecting any given element earlier helps more than selecting it later. 
Given a set of constraints on the desired summary, and a (pre-designed or learned) submodular utility function $f$ that quantifies the representativeness $f(S)$ of a subset $S$ of items, 
data summarization  can  be naturally reduced to a constrained submodular optimization problem.

In this paper, we are motivated by applications of \emph{non-monotone} submodular maximization. 
In particular, we consider video summarization in a streaming setting, where video frames are produced at a fast pace, and we want to keep an updated summary of the video so far, with little or no memory overhead. This has important applications e.g. in surveillance cameras, wearable cameras, and astro video cameras,
which generate data at too rapid a pace to efficiently analyze and store it in main memory. The same framework can be applied more generally in many settings where we need to extract a small subset of data from a large stream to train or update a machine learning model.
At the same time, various constraints may be imposed by the underlying summarization application. These may range from a simple limit on the size of the summary to more complex restrictions such as focusing on particular individuals or objects, or excluding them from the summary. These requirements often arise in real-world scenarios to consider privacy (e.g. in case of surveillance cameras) or personalization (according to users' interests).

In machine learning, Determinantal Point Processes (DPP) have been proposed as computationally efficient methods for selecting a diverse subset from a ground set of items \cite{kulesza2012determinantal}. They have recently shown great success for video summarization \cite{gong2014diverse}, %as well as problems like 
document summarization \cite{kulesza2012determinantal} and information retrieval \cite{gillenwater2012discovering}. While finding the most likely configuration (MAP) is NP-hard, the DPP probability is a log-submodular function, and submodular optimization techniques can be used to find a near-optimal solution. 
In general, the above submodular function is very non-monotone, and we need techniques for maximizing a non-monotone submodular function in the streaming setting.
Although efficient streaming methods have been recently developed for maximizing a monotone submodular function $f$ with a variety of constraints, there is no effective streaming solution for non-monotone submodular maximization under general types of constraints. %in the streaming setting.

In this work, we provide \SLS, the first single pass streaming algorithm for non-monotone submodular function maximization, subject to the intersection of a %$p$-system
collection of independence systems $\ind$ and $d$ knapsack constraints.
Our approach builds on local search, a %powerful and 
widely used technique for maximizing non-monotone submodular functions in a batch mode. 
Local search, however, needs multiple passes over the input, and hence does not directly extend to the streaming setting, where we are only allowed to make a single pass over the data. 
This work provides a general framework within which we can use any streaming monotone submodular maximization algorithm, \pstream, with approximation guarantee $\alpha$ under a collection of independence systems $\ind$. For any such monotone algorithm, \SLS provides a constant %$(1-\alpha)/(2/\alpha+2d-1)$
$1/\big(1\!+\!2/\sqrt{\alpha}\!+\!1/\alpha +2d(1\!+\!\sqrt{\alpha})\big)$
approximation guarantee for maximizing a non-monotone submodular function under the intersection of $\ind$ and $d$ knapsack constraints. Furthermore, \SLS needs a memory and update time that is larger than \pstream with a factor of $O(\log(k)/ \sqrt{\alpha})$, where $k$ is the size of the largest feasible solution. 
Using parallel computation, the increase in the update time can be reduced to $O(1/\sqrt{\alpha})$, making our approach an appealing solution in real-time scenarios.
We show that for video summarization, our algorithm leads to streaming solutions that provide competitive utility when compared with those obtained via centralized methods, at a small fraction of the computational cost, i.e. more than 1700 times faster.

\section{Related Work}
Video summarization aims to retain diverse and representative
frames according to criteria such as representativeness, diversity, interestingness, or frame importance \cite{ngo2003automatic,liu2006optimization,lee2012discovering}. 
This often requires hand-crafting to combine the criteria effectively.
Recently, \citet{gong2014diverse} proposed a supervised subset selection method using DPPs.
Despite its superior performance, this method uses an exhaustive search for MAP inference, which makes it inapplicable for producing real-time summaries.

Local search has been widely used for submodular maximization subject to various constraints. This includes the analysis of greedy and local search by %Nemhauser et al. 
\citet{nemhauser1978analysis} providing a $1/(p+1)$ approximation guarantee for monotone submodular maximization under $p$ matroid constraints.
For non-monotone submodular maximization, the most recent results include a $(1+O(1/\sqrt{p}))p$-approximation subject to a $p$-system constraints \cite{feldman2017greed}, a $1/5-\eps$ approximation under $d$ knapsack constraints \cite{lee2009non}, and a $(p+1)(2p + 2d + 1)/p$-approximation for maximizing a general submodular function subject to a $p$-system and $d$ knapsack constraints \cite{mirzasoleiman2016fast}.

Streaming algorithms for submodular maximization have gained increasing attention for producing online summaries. For monotone submodular maximization, %Badanidiyuru et al. 
\citet{badanidiyuru2014streaming} proposed a single pass  algorithm with a $1/2 \!-\! \epsilon$ approximation guarantee under a cardinality constraint $k$, using $O(k \log k /\epsilon)$ memory.  
Later, %Chakrabarti and Kale
\citet{chakrabarti2015submodular} provided a $1/4p$ approximation guarantee for the same problem under the intersection of $p$ matroid constraints.
However, the required memory 
increases  polylogarithmically with the size of the data. 
Finally, %Chekuri et al. 
\citet{chekuri2015streaming} presented deterministic and randomized algorithms for maximizing monotone and non-monotone submodular functions subject to a broader range of constraints, namely a $p$-matchoid. 
For maximizing a monotone submodular function, their proposed method gives a $1/4p$ approximation using $O(k \log k / \epsilon^2)$ memory ($k$ is the size of the largest feasible solution). For non-monotone functions, they provide a deterministic $1/(9p+1)$ approximation using the $1\!/(p\!+\!1)$ offline approximation of \citet{nemhauser1978analysis}. %under a $p$-matchoid constraint. 
Their randomized algorithm provides a $1/(4p+1/\tau_p)$ approximation in expectation, where $\tau_p=(1-\eps)(2-o(1))/(ep)$ \cite{feldman2011unified} is the 
%approximation guarantee for maximizing a non-negative submodular function in the offline setting.
offline approximation for maximizing a non-negative submodular function.

Using the monotone streaming algorithm of \citet{chekuri2015streaming} with $1/4p$ approximation guarantee, our framework provides a $1/(4p+4\sqrt{p}+1)$ approximation %guarantee 
for maximizing a non-monotone function under a $p$-matchoid constraint, which is a significant improvement over the work of \citet{chekuri2015streaming}. 
Note that any monotone streaming algorithm with approximation guarantee under a set of independence systems $\ind$ (including a $p$-system constraint, once such an algorithm exists) can be integrated into our framework to provide approximations %guarantees 
for non-monotone submodular maximization under the same set of independence systems $\ind$, \emph{and $d$ knapsack constraints}.

\section{Problem Statement}
We consider the problem of summarizing a stream of data by selecting, on the fly, a subset that maximizes a utility function $f : 2^V \rightarrow \mathbb{R}_+$. The utility function is defined on $2^V$ (all subsets of the entire stream $V$), and for each $S \subseteq V$, $f(S)$ quantifies how well $S$ represents the ground set $V$.
We assume that $f$ is \emph{submodular}, a property that holds for many widely used such utility functions. This means that for any  two sets $S\subseteq T\subseteq V$ and any element $e\in V\setminus T$ we have %that 
$$f(S\cup \{e\})-f(S)\geq f(T\cup \{e\})-f(T).$$
We denote the \emph{marginal gain} of adding an element $e \in V$ to a summary $S \subset V$ by $f_S(e) = f(S\cup\{e\})-f(S)$. The function $f$ is \emph{monotone} if $f_S(e)\geq 0$ for all $S\subseteq V$.
Here, we allow $f$ to be non-monotone.
Many data summarization applications can be cast as an instance of constrained submodular maximization under a set $\zeta \subseteq 2^V$ of constraints:
$$S^* = \argmax_{S \in \zeta}f(S).$$  
In this work, we consider a collection of independence systems and multiple knapsack constraints. An independence system is a pair $\cM^I=(V,\mathcal{I})$ where $V$ is a finite (ground) set, and $\mathcal{I} \subseteq 2^V$ is a family of independent subsets of $V$ 
satisfying the following two properties. (i) $\emptyset \in \mathcal{I}$, and (ii) for any $A \! \subseteq \!B\! \subseteq \!V$, \!$B\! \in \!\mathcal{I}$ implies that $A \in \mathcal{I}$ (hereditary property).
A \textit{matroid} $\mathcal{M}=(V,\mathcal{I})$ is an independence system with exchange property: if $A, B \in \mathcal{I}$ and $|B| > |A|$, there is an element $e \in B \setminus A$ such that $A \cup \{e\} \in \mathcal{I}$. 
The maximal independent sets of $\mathcal{M}$ share a common cardinality, called the rank of $\mathcal{M}$.
A \textit{uniform} matroid is the family of all subsets of size at most $l$. In a \textit{partition} matroid, we have a collection of disjoint sets $B_i$ and integers $0\!\leq l_i\!\leq |B_i|$ where a set $A$ is independent if for every index $i$, we have $|A\cap B_i|\leq l_i.$ 
A \textit{$p$-matchoid} generalizes matchings and intersection of matroids. For $q$ matroids $\cM_\ell=(V_{\ell}, \mathcal{I}_{\ell})$, $\ell \in [q]$, defined over overlapping ground sets $V_\ell$, 
and for $V= \cup_{\ell=1}^q V_{\ell}$, $\mathcal{I}=\{S \subseteq V: S \cap V_\ell \in \mathcal{I}_\ell ~~\forall \ell \}$, we have that $\cM^p=(V, \mathcal{I})$ is a $p$-matchoid if
%it requires that 
every element $e \in V$ is a member of $V_\ell$ for at most $p$ indices.
Finally, a \textit{$p$-system} is the most general type of constraint we consider in this paper. It requires that  if $A,B\in \cI$  are two maximal sets, then $|A| \leq p|B|$.  
A \textit{knapsack} constraint is defined by a cost function $c:V\rightarrow \mathbb{R}_+$. A set $S\subseteq V$ is said to satisfy the knapsack constraint if $c(S)=\sum_{e\in S}c(e)\leq W$. Without loss of generality, we assume $W=1$ throughout the paper.

The goal in this paper is to maximize a (non-monotone) submodular function $f$ subject to a set of constraints $\zeta$ defined by the intersection of a collection of independence systems $\ind$, and $d$ knapsacks. In other words, we would like to find a set $S\in \cI$ that maximizes $f$ where for each set of knapsack costs $c_i, i \in [d]$, we have $\sum_{e\in S} c_i(e)\leq 1$. 
We assume that the ground set $V = \{e_1, \cdots , e_n\}$ is received from the stream in some arbitrary order.
At each point $t$ in time, the algorithm may maintain a memory $M_t \!\subset \!V$ of points, and must be ready to output a %candidate 
feasible solution $S_t \subseteq M_t$, such that $S_t \in \zeta$. 
%%%Upon receiving an element $e_t$ from the stream, the algorithm may elect to 1) insert it into its memory, 2) discard some elements in its memory and accept $e_t$ instead, or 3) discard $e_t$.

\vspace{-1mm}
\section{Video Summarization with DPPs}\label{sec:dpp}
Suppose that we are receiving a stream of video frames, e.g. from a surveillance or a wearable camera, and we wish to select a subset of frames that concisely represents all the diversity contained in the video. 
Determinantal Point Processes (DPPs) are good tools for modeling diversity in such applications.
DPPs \cite{macchi1975coincidence} are distributions over subsets with a preference for diversity.
Formally, a DPP $\mathcal{P}$ on a set of items $V=\{1,2,...,N\}$ defines a discrete probability distribution on $2^V$\!\!, such that the probability of every
% the probability of observing subset
 $S\!\subseteq \!V$ is 
\begin{equation}\label{eq:dpp}
\mathcal{P}(Y=S) = \frac{\det (L_S)}{\det(I + L)},
\end{equation}
where $L$ is a positive semidefinite kernel matrix, and $L_S \equiv [L_{ij}]_{i,j \in S}$, is the restriction of $L$ to the entries indexed by elements of $S$, 
and $I$ is the $N \times N$ identity matrix. 
In order to find the most diverse and informative feasible subset, we need to solve the NP-hard problem of finding $\arg \max_{S\in \mathcal{I}} \det(L_S)$ \cite{ko1995exact}, 
where $\mathcal{I} \subset 2^V$ is a given family of feasible solutions.  
However, the logarithm $f(S)=\log \det(L_S )$ is a (non-monotone) submodular function \cite{kulesza2012determinantal}, and we can apply submodular maximization techniques. 

Various constraints can be imposed while maximizing the above non-monotone submodular function. %utility function. 
In its simplest form, we can partition the video into $T$ segments, and define a diversity-reinforcing  partition matroid to select at most $k$ frames from each segment. Alternatively, various content-based constraints can be applied, e.g., we can use object recognition to select at most $k_i \geq 0$ frames showing person $i$, %in the video, 
or to find a summary that is focused on a particular person or object. 
Finally, each frame can be associated with multiple costs, based on qualitative factors such as resolution, contrast, luminance, or the probability that the given frame contains an object.
Multiple knapsack constraints, one for each quality factor, can then limit the total costs of the elements of the solution and enable us to produce a summary closer to human-created summaries by filtering uninformative frames.

\section{Streaming algorithm for constrained submodular maximization}
In this section, we describe our streaming algorithm for maximizing a non-monotone submodular function subject to the intersection of a collection of independence systems and $d$ knapsack constraints. 
Our approach builds on local search, a %powerful and
widely used technique for maximizing non-monotone submodular functions. It starts from a candidate solution $S$ and iteratively increases the value of the solution by either including a new element in $S\!$ or discarding one of the elements of $S$ \cite{feige2011maximizing}.
\citet{gupta2010constrained}  showed that similar results can be obtained with much lower complexity by using algorithms for \emph{monotone} submodular maximization, which, however, are run multiple times.
Despite their effectiveness, these algorithms need multiple passes over the input and do not directly extend to the streaming setting, where we are only allowed to make one pass over the data. In the sequel, we show how local search can be implemented in a single pass in the streaming setting.

\subsection{\SLS for a collection of independence systems}
The simple yet crucial observation underlying
the approach of \citet{gupta2010constrained} is the following. The solution obtained by approximation algorithms for monotone submodular functions often satisfy $f(S) \geq \alpha f(S \cup C^*)$, where $1 \geq \alpha > 0$, and $C^*$ is the optimal solution. In the monotone case $f(S \cup C^*) \geq f(C^*)$, and we obtain the desired approximation factor $f(S) \geq \alpha f(C^*)$. However, this does not hold for non-monotone functions. But, if $f(S \cap C^*)$ provides a good fraction of the optimal solution, then we can find
a near-optimal solution for \emph{non-monotone} functions even from the result of an algorithm for \emph{monotone} functions, by pruning elements in $S$ using unconstrained maximization. This still retains a feasible set, since the constraints are downward closed. Otherwise,  if $f(S \cap C^*) \leq \eps \opt$, then running another round of the algorithm on the remainder of the ground set will lead to a good solution.

\begin{algorithm}[!htb]%[H]
	\caption{\SLS for independence systems}\label{alg:psys}
	\begin{algorithmic}[1]
		\INPUT $f:2^V \rightarrow \RR_+$, a membership oracle for independence systems $\cI \subset 2^V$; and a monotone streaming algorithm \pstream with $\alpha$-approximation under $\cI$.  
		\OUTPUT A set $S \subseteq V$ satisfying $S \in \cI$. 
		\WHILE {stream is not empty}
		\STATE $D_0 \leftarrow \{e\}$
		~~~ \(\triangleright\) $e$ is the next element from the stream.
		\STATE \(\triangleright\) \LS iterations
		\FOR {$i = 1$ to $\lceil 1/\sqrt{\alpha} + 1\rceil$} 
		\STATE \(\triangleright\) $D_i$ is the discarded set by $\pstream_i$
		\STATE
		$[D_{i},S_{i}]\!\!=\pstream_i$($D_{i-1}$) 
		\STATE $S'_i =$\textsc{Unconstrained-Max}($S_i$).
		\ENDFOR
		\STATE $S = \argmax_i \{f(S_i),f(S'_i)\}$
		\ENDWHILE
		\STATE Return $S$ %$\argmax\{f(S)|S\in U\}$
	\end{algorithmic}
\end{algorithm}
\vspace{-1mm}

Backed by the above intuition, we aim to build multiple disjoint solutions simultaneously within a single pass over the data.
Let \pstream be a single pass streaming algorithm for monotone submodular maximization under a collection of independence systems, with approximation factor $\alpha$.
Upon receiving a new element from the stream, \pstream can choose (1) to insert it into its memory, (2) to replace one or a subset of elements in the memory by it, or otherwise (3) the element gets discarded forever. %and cannot be used later by the algorithm. 
The key insight for our approach is that it is possible to  build other solutions from the elements discarded by \pstream. 
Consider a chain of $q\!=\!\lceil 1/\sqrt{\alpha}\!+\!1\rceil$ instances of our streaming algorithm, i.e. $\{\pstream_1, \cdots, \pstream_q \}$. 
Any element $e$ received from the stream is first passed to $\pstream_1$. If $\pstream_1$ discards $e$, or adds $e$ to its solution and instead discards a set $D_1$ of elements from its memory, then we pass the set $D_1$ of discarded elements on to be processed by $\pstream_2$.
Similarly, if a set of elements $D_2$ is discarded by $\pstream_2$, we pass it to $\pstream_3$, and so on. 
The elements discarded by the last instance $\pstream_q$ are discarded forever. 
%Finally, 
At any point in time that we want to return the final solution, we run unconstrained submodular maximization (e.g. the algorithm of \citet{buchbinder2015tight}) on each solution $S_i$ obtained by $\pstream_i$ to get $S'_i$, and return the best solution among $\{S_i, S'_i\}$ for $i \in [1,q]$.

\begin{theorem}\label{thm:psys}
	Let \pstream be a streaming algorithm for monotone submodular maximization under a collection of independence systems $\cI$ with approximation guarantee $\alpha$.
	%	For any set $C\in \cI$,  
	Alg. \ref{alg:psys} returns a set $S\in \cI$ with
	\begin{equation*}
	f(S) \geq \frac{1}{(1+1/\sqrt{\alpha})^2}\opt,
	\end{equation*}
	using memory $O(M/\sqrt{\alpha})$, and average update time $O(T /\sqrt{\alpha})$ per element, where $M$ and $T$ are the memory and update time of \pstream. 
\end{theorem}
The proof of all the theorems can be found in 
\cite{mirzasoleiman2017streaming}.

\begin{algorithm}[!htb]%[H]
	\caption{\SLS for independence systems $\cI$ and $d$ knapsacks}
	\label{alg:pknapsack}
	\begin{algorithmic}[1]
		\INPUT $f:2^V \rightarrow \RR_+$, a membership oracle for independence systems $\cI \subset 2^V$; $d$ knapsack-cost functions
		$c_j:V \rightarrow [0,1]$; \pstream; and an upper bound $k$ on the cardinality of the largest feasible solution.
		\OUTPUT A set $S \subseteq V$ satisfying $S \in \cI$ and $c_j(S) \leq 1 ~ \forall j$.
		%\STATE $U=\{\}.$
		\STATE $m = 0.$
		\WHILE {stream is not empty}
		\STATE $D_0 \leftarrow \{e\}$
		~~~ \(\triangleright\) $e$ is the next element from the stream.
		\STATE $m = \max(m,f({e})), ~e_m = \argmax_{e\in V}f(e)$.
		\STATE
		$\gamma = \frac{2\cdot  m}{(1+1/\sqrt{\alpha})(1+1/\sqrt{\alpha}+2d\sqrt{\alpha})}$
		\STATE $R=\left\{\gamma, (1+\epsilon)\gamma, (1+\epsilon)^2\gamma, (1+\epsilon)^3\gamma, \ldots ,\gamma k\right\}$
		\FOR {$\rho \in R$ in parallel}
		\STATE \(\triangleright\) \LS
		\FOR {$i = 1$ to $\lceil 1/\sqrt{\alpha} + 1\rceil$}
		\STATE \(\triangleright\) picks elements only if  $\frac{f_{S_i}(e)}{\sum_{j=1}^d c_{je}}\geq \rho$
		\STATE $[D_{i},S_{i}]\!\!=\pdstream_i$($D_{i-1}, \rho$)
		\STATE \(\triangleright\) unconstrained submodular maximization
		\STATE $S'_i =$\textsc{Unconstrained-Max}($S_i$).
		\ENDFOR
		\STATE $S_\rho = \argmax_i \{f(S_i),f(S'_i)\}$
		\ENDFOR
		\STATE $S = \argmax_{\rho \in R} f(S_\rho)$
		\ENDWHILE
		\STATE Return $\argmax\{f(S), f(\{e_m\})$ 
	\end{algorithmic}
\end{algorithm}

We make Theorem~\ref{thm:psys} concrete via an example:  \citet{chekuri2015streaming} proposed a $1/4p$-approximation streaming algorithm for monotone submodular maximization under a $p$-matchoid constraint. 
Using this algorithm as \pstream in  \SLS, we obtain: %obtain the following result:

\begin{corollary}\label{col:psys}
	With \textsc{Streaming Greedy} of \citet{chekuri2015streaming} as \pstream, \SLS yields a solution $S\in \cI$ with approximation guarantee ${1}/(1+{2\sqrt{p}})^2$, 
	using $O(\sqrt{p}k \log(k)/\eps)$ memory and $O(p\sqrt{p} k\log(k)/\eps)$ average update time per element, where $\mathcal{I}$ are the independent sets of a $p$-matchoid, and $k$ is the size of the largest feasible solution.
\end{corollary}

\begin{table*}[!htb]
	\caption{Performance of various video summarization methods with segment size 10 on YouTube and OVP datasets, measured by F-Score (F), Precision (P), and Recall (R).
	}\label{table:scores}
	\begin{center}
		\begin{tabular}{|c c p{1.5cm} p{2cm} p{1.5cm} p{2.4cm} c c c|}
			\hline
			&\multirow{1}{*}{} & \multicolumn{2}{c}{Alg. of \cite{gong2014diverse}$^\text{(centralized)}$}& \multicolumn{2}{c}{\AlgF %\cite{mirzasoleiman2016fast}
				$^\text{(centralized)}$} & \multicolumn{2}{c}{\textsc{\SLS}}& \\\cline{3-8}			
			&& Linear &  N. Nets & Linear &  N. Nets& Linear &  N. Nets&\\
			\hline %\hline
			\multicolumn{1}{|c|}{\multirow{3}{*}{\begin{turn}{0}YouTube\end{turn}} }&\multicolumn{1}{c|}{F}&
			57.8$\pm$0.5& 60.3$\pm$0.5&57.7$\pm$0.5&60.3$\pm$0.5&58.3$\pm$0.5&59.8$\pm$0.5 &\\
			\multicolumn{1}{|c|}{}&\multicolumn{1}{c|}{P}&
			54.2$\pm$0.7&59.4$\pm$0.6&54.1$\pm$0.5&59.1$\pm$0.6&55.2$\pm$0.5&58.6$\pm$0.6&\\
			\multicolumn{1}{|c|}{}&\multicolumn{1}{c|}{R}& 69.8$\pm$0.5&64.9$\pm$0.5&70.1$\pm$0.5&64.7$\pm$0.5&70.1$\pm$0.5&64.2$\pm$0.5&\\
			\hline 
			\multicolumn{1}{|c|}{\multirow{3}{*}{\begin{turn}{0}OVP\end{turn}} }&\multicolumn{1}{c|}{F}&
			75.5$\pm$0.4&77.7$\pm$0.4&75.5$\pm$0.3&78.0$\pm$0.5&74.6$\pm$0.2&75.6$\pm$0.5&\\
			\multicolumn{1}{|c|}{}&\multicolumn{1}{c|}{P}&
			77.5$\pm$0.5&75.0$\pm$0.5&77.4$\pm$0.3&75.1$\pm$0.7&76.7$\pm$0.2&71.8$\pm$0.7 &\\
			\multicolumn{1}{|c|}{}&\multicolumn{1}{c|}{R}&
			78.4$\pm$0.5&87.2$\pm$0.3&78.4$\pm$0.3&88.6$\pm$0.2&76.5$\pm$0.3&86.5$\pm$0.2 &\\ 
			\hline  
		\end{tabular}
	\end{center}
\end{table*}

Note that any monotone streaming algorithm with approximation guarantee $\alpha$ under a collection of independence systems $\cI$ can be integrated into Alg. \ref{alg:psys} to provide approximation guarantees for non-monotone submodular maximization under the same set $\cI$ of constraints. For example, as soon as there is a subroutine for monotone streaming submodular maximization under a $p$-system in the literature, one can use it in Alg. \ref{alg:psys} as \pstream, and get the guarantee provided in Theorem \ref{thm:psys} for maximizing a non-monotone submodular function under a $p$-system, in the streaming setting.

\subsection{\SLS for independence systems and multiple knapsack constraints}

To respect multiple knapsack constraints in addition to the collection of independence systems $\cI$, we integrate the idea of a density threshold \cite{sviridenko2004note} 	%badanidiyuru2014fast,
into our local search algorithm. 
We use a (fixed) density threshold $\rho$ to restrict the \pstream algorithm to only pick elements if the function value per unit size of the selected elements is above the given threshold. We call this new algorithm \pdstream. The threshold should be carefully chosen to be below the value/size ratio of the optimal solution. To do so, we need to know (a good approximation to) the value of the optimal solution \opt. To obtain a rough estimate of \opt, it suffices to know the maximum value $m = \max_{e\in V} f({e})$ of any singleton element: submodularity implies that $m\!\leq\!\opt\!\leq\!km$, where $k$ is an upper bound on the cardinality of the largest feasible solution satisfying all constraints.  We update the value of the maximum singleton element on the fly \cite{badanidiyuru2014streaming}, and lazily instantiate the thresholds to $\log (k) / \epsilon$ different possible values $(1+\epsilon)^i \in [\gamma, \gamma k]$, for $\gamma$ defined in Alg. \ref{alg:pknapsack}. We show that for at least one of the discretized density thresholds we obtain a good enough solution. 

\begin{theorem}\label{thm:pknapsack}
	\SLS (outlined in Alg.~\ref{alg:pknapsack}) guarantees %has an approximation guarantee
	\begin{align*}
	f(S) \geq \frac{1-\epsilon}{(1+1/\sqrt{\alpha})(1+2d\sqrt{\alpha}+1/\sqrt{\alpha})}\opt,
	\end{align*}
	with memory $O(M \log(k)/(\epsilon \sqrt{\alpha}))$, and average update time $O(T \log(k)/(\epsilon \sqrt{\alpha}))$ per element, where $k$ is an upper bound on the size of the largest feasible solution, and $M$ and $T$ are the memory and update time of the \pstream algorithm.
\end{theorem}

\begin{corollary}\label{col:knapsack}
	By using \textsc{Streaming Greedy} of \citet{chekuri2015streaming}, we get that \SLS has an approximation ratio $(1+\eps)(1+4p+4\sqrt{p}+d(2+1/\sqrt{p}))$ with $O(\sqrt{p}k \log^2(k)/\eps^2)$ memory and update time $O(p\sqrt{p} k \log^2(k)/\epsilon^2)$ per element, where $\mathcal{I}$ are the independent sets of the $p$-matchoid constraint, and $k$ is the size of the largest feasible solution.
\end{corollary}

\subsubsection{Beyond the Black-Box.}
Although the DPP probability in Eq. \ref{eq:dpp} only depends on the selected subset $S$, in many applications $f(S)$ may depend on the entire data set $V$.
So far, we have adopted the common assumption that $f$ is given in terms of a value oracle (a black box) that computes $f(S)$. Although in practical settings this assumption might be violated, many objective functions are \textit{additively decomposable} over the ground set $V$ \cite{mirzasoleiman2016distributed}. That means,  $f(S)=\frac{1}{V}\sum_{e \in V}f_e(S)$, where $f_e(S)$ is a non-negative submodular function associated with every data point $e \in V$, and $f_e(.)$ can be evaluated without access to the full set $V$. For decomposable functions, we can approximate $f(S)$ by $f_W(S)=\frac{1}{W}\sum_{e \in W}f_e(S)$, where $W$ is a uniform sample from the stream (e.g. using reservoir sampling \cite{vitter1985random}). 

\begin{theorem}[\textbf{{\citet{badanidiyuru2014streaming}}}]
	Assume that $f$ is decomposable, all of $f_e(S)$ are bounded, and w.l.o.g. $|f_e(S)| \!\leq \!1$. Let $W$ be uniformly sampled from $V$\!. Then for $|W| \geq \frac{2 k^2\log(2/\delta)+2k^3 \log(V)}{\eps^2}$, we can ensure that with probability $1\!-\!\delta$, \SLS guarantees %has an approximation guarantee
	$$f(S)\geq \frac{1-\epsilon}{(1+1/\sqrt{\alpha})(1+2d\sqrt{\alpha}+1/\sqrt{\alpha})}(\opt-\eps).$$
\end{theorem}

\section{Experiments}
In this section, we apply \SLS to  
video summarization in the streaming setting.
The main goal of this section is to validate our theoretical results and demonstrate the effectiveness of our method in practical scenarios, where the existing streaming algorithms are incapable of providing any  quality guarantee for the solutions.
In particular, for streaming non-monotone submodular maximization under a collection of independence systems and multiple knapsack constraints, none of the previous works provide any theoretical guarantees. 
We use the streaming algorithm of \citet{chekuri2015streaming} for monotone submodular maximization under a $p$-matchoid constraint as \pstream, and
compare the performance of our method\footnote{Our code is available at %https://github.com/baharanm/non-monotone-stream
github.com/baharanm/non-mon-stream} with exhaustive search \cite{gong2014diverse}, and a centralized method for maximizing a non-monotone submodular function under a $p$-system and multiple knapsack constraints, \AlgF \cite{mirzasoleiman2016fast}.

\begin{figure*}[!htb]
	\centering  
	\subfloat[YouTube Linear\label{subfig:you-fs}]{
		\includegraphics[width=.278\textwidth]{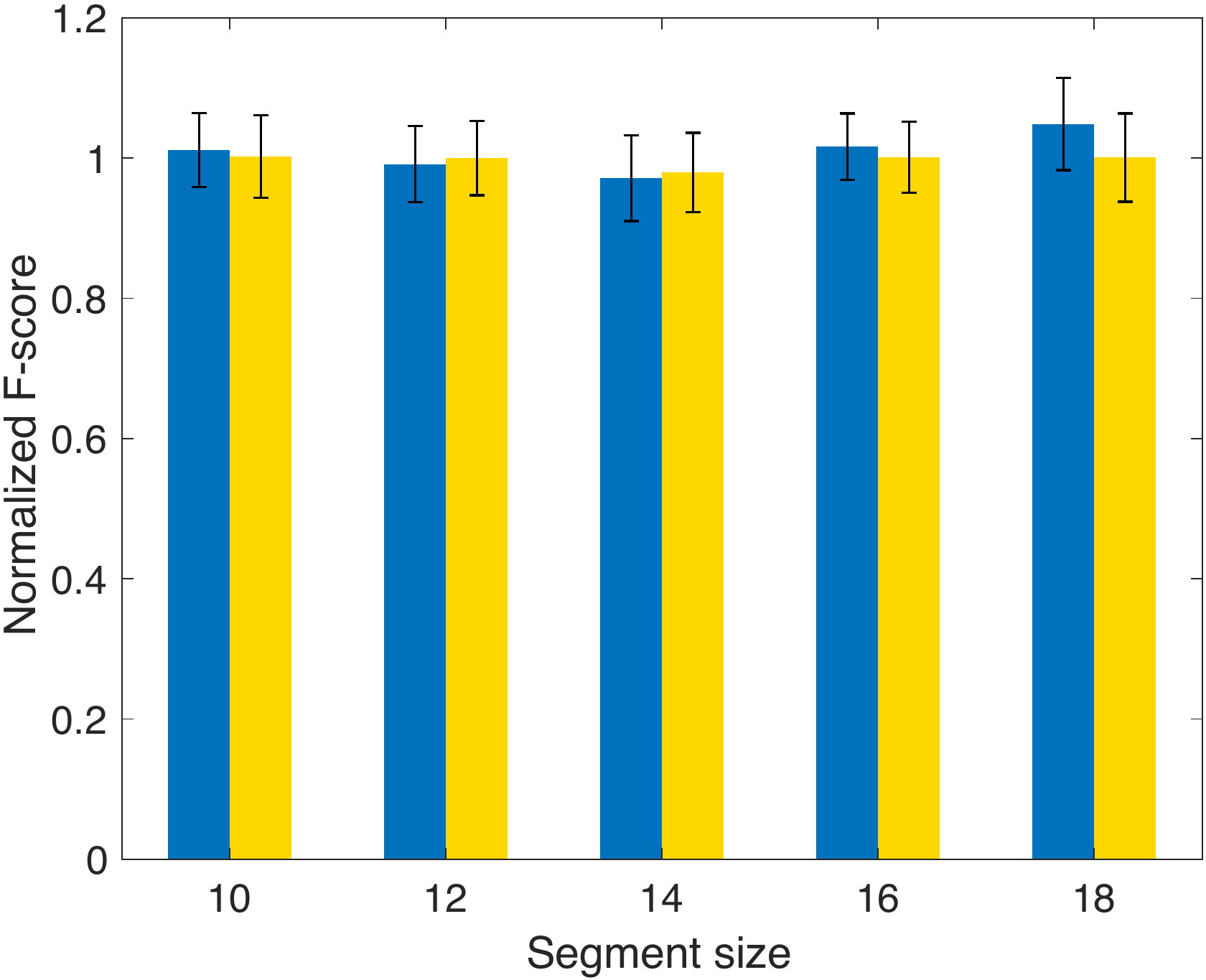}}\hspace{3mm}
	\subfloat[YouTube Linear\label{subfig:you-speed}]{
		\includegraphics[width=.283\textwidth]{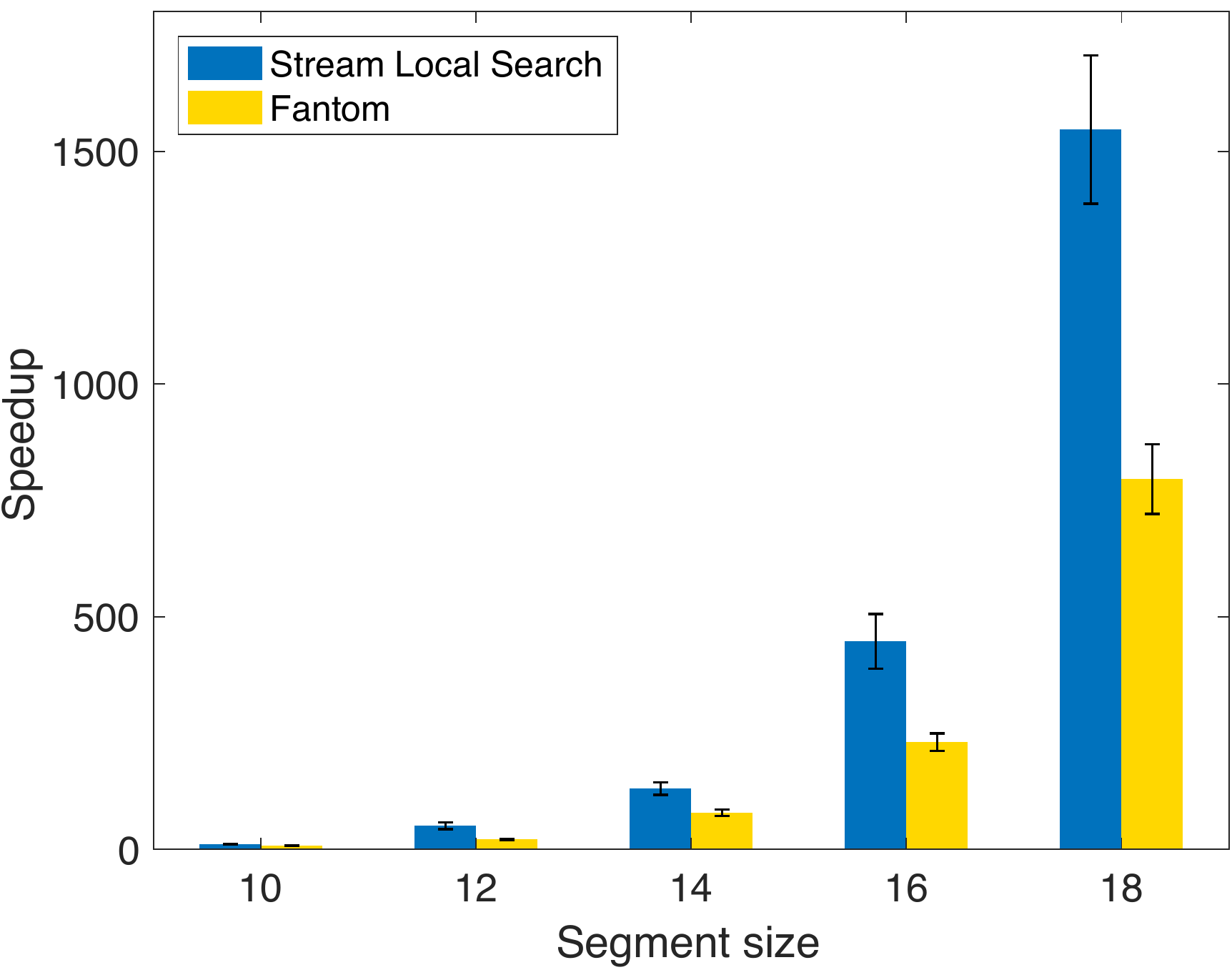}}\hspace{5mm}
	\subfloat[YouTube Linear\label{subfig:you-linear}]{
		\includegraphics[width=.267\textwidth]{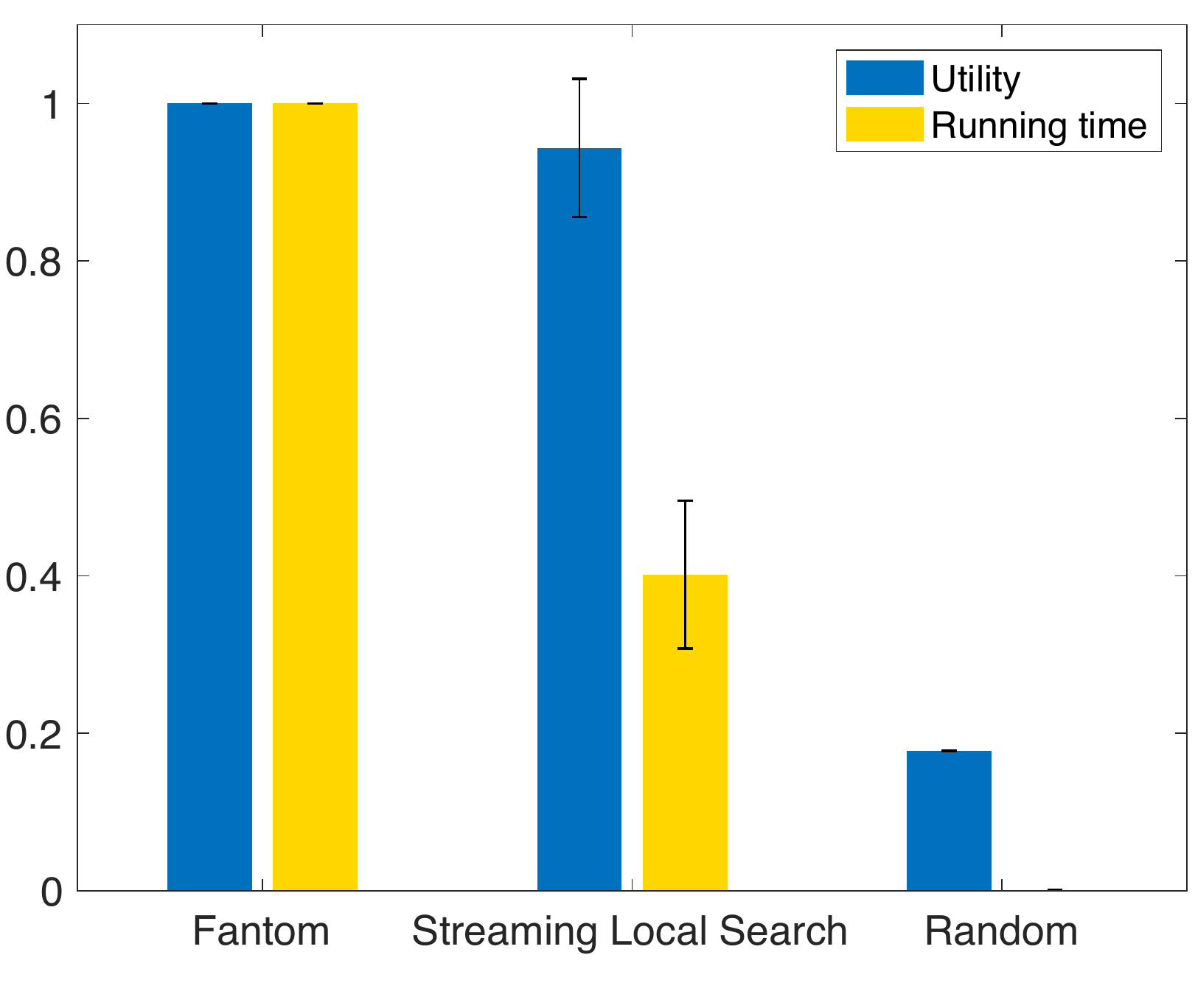}}
	\\
	\subfloat[YouTube N. Nets\label{subfig:you-nn-fs}]{
		\includegraphics[width=.278\textwidth]{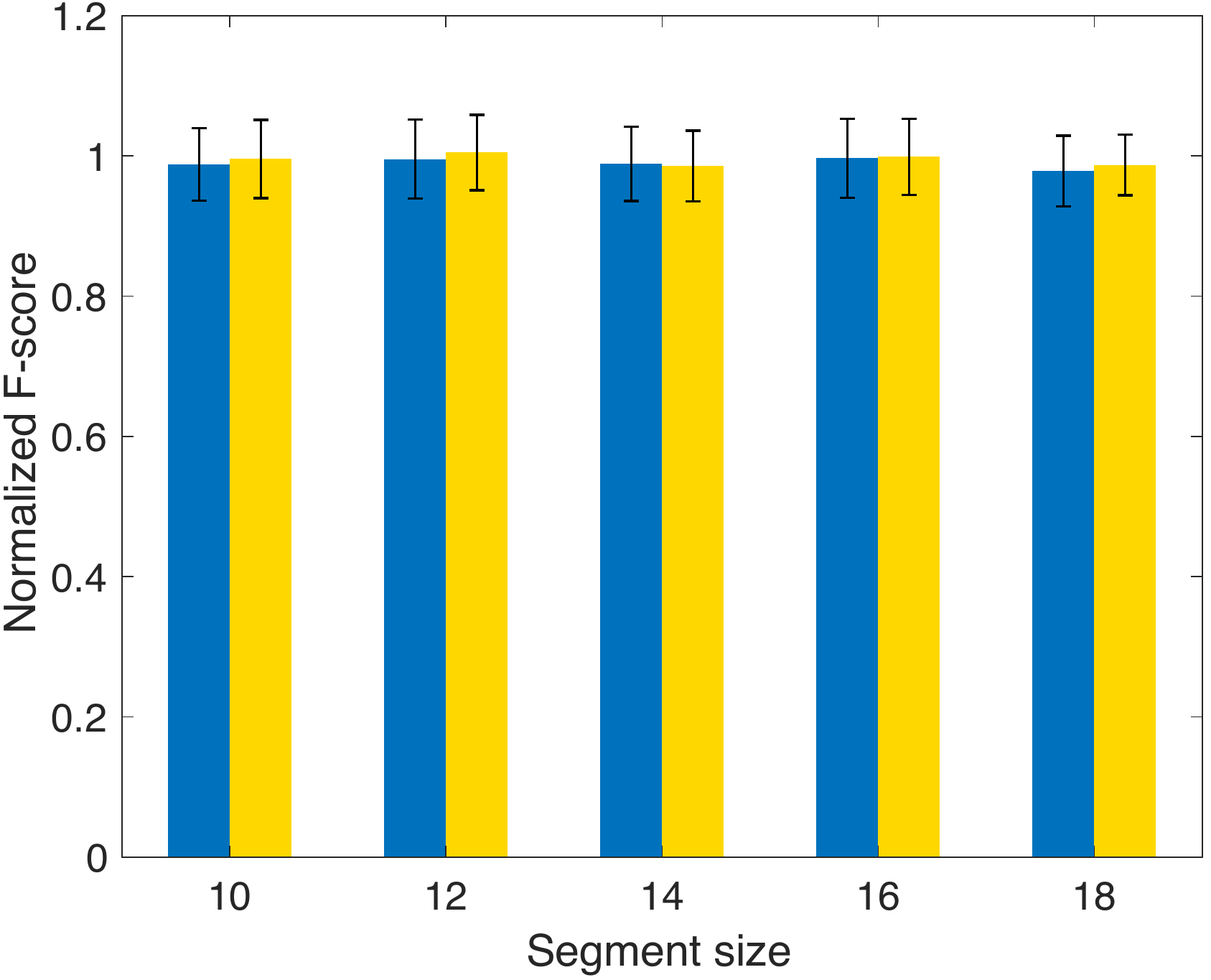}}\hspace{3mm}
	\subfloat[YouTube N. Nets\label{subfig:you-nn-speed}]{
		\includegraphics[width=.283\textwidth]{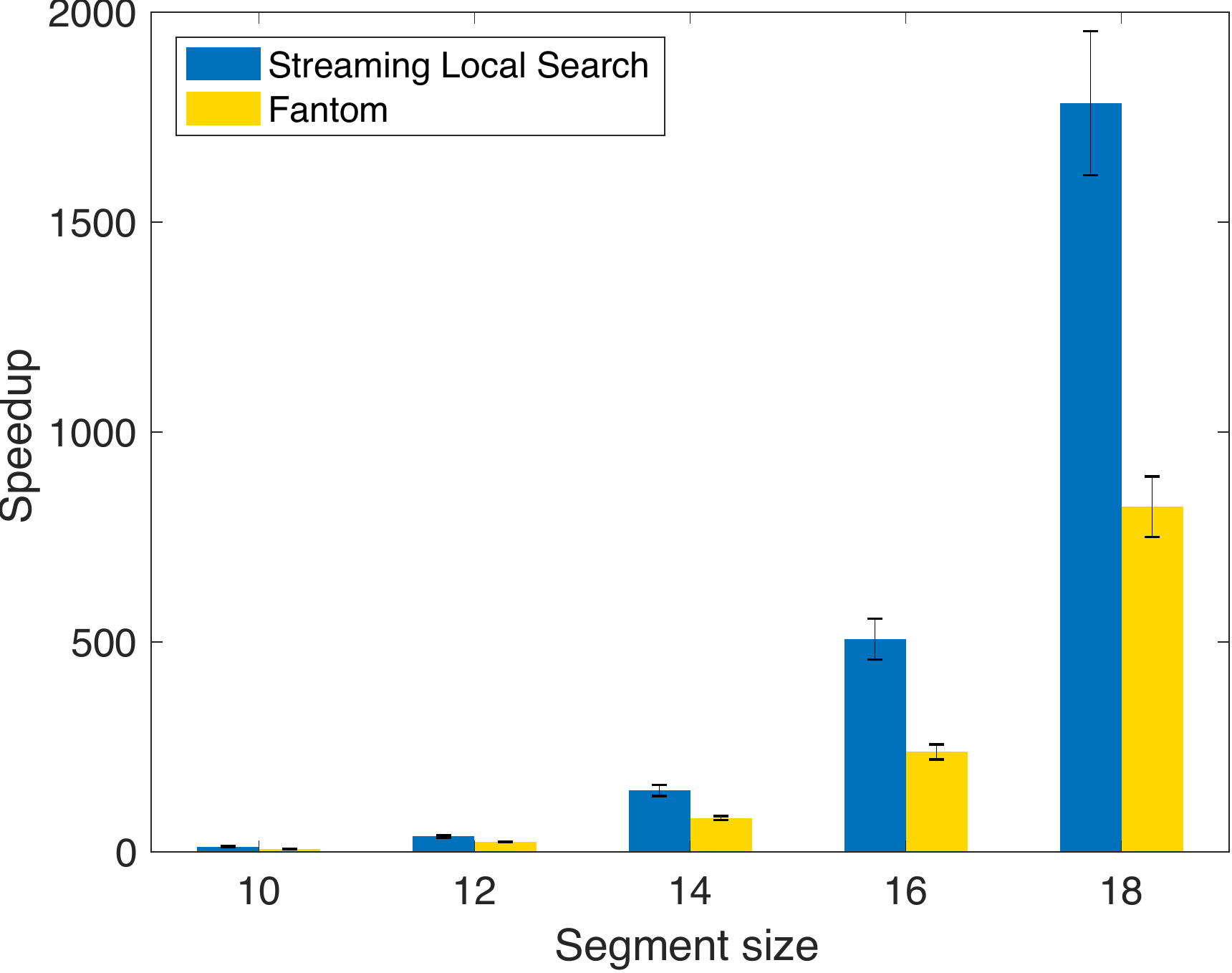}\hspace{5mm}
	}
	\subfloat[YouTube N. Nets\label{subfig:you-nn}]{
		\includegraphics[width=.267\textwidth]{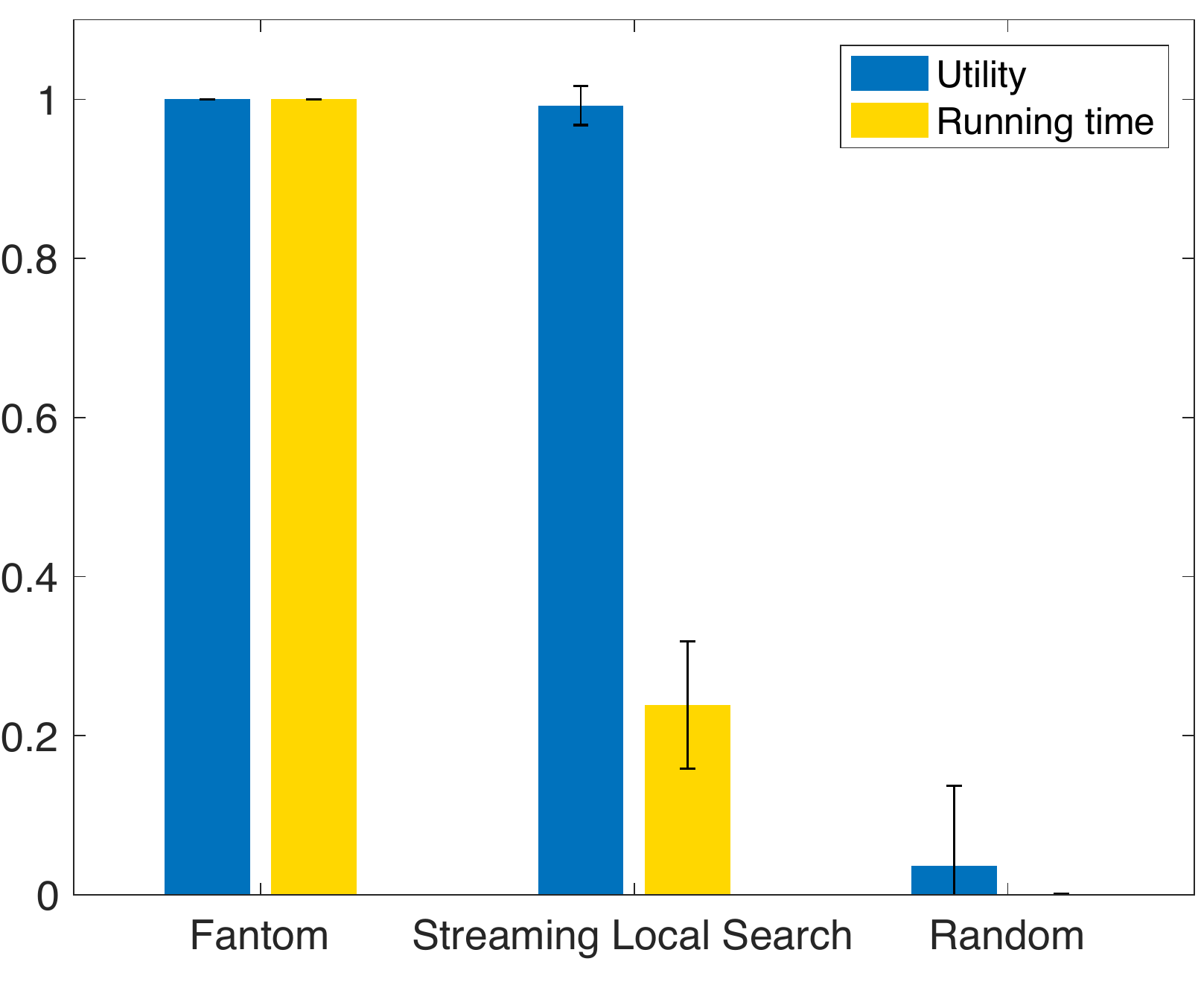}}
	\\
	\subfloat[OVP Linear\label{subfig:ovp-fs}]{
		\includegraphics[width=.278\textwidth]{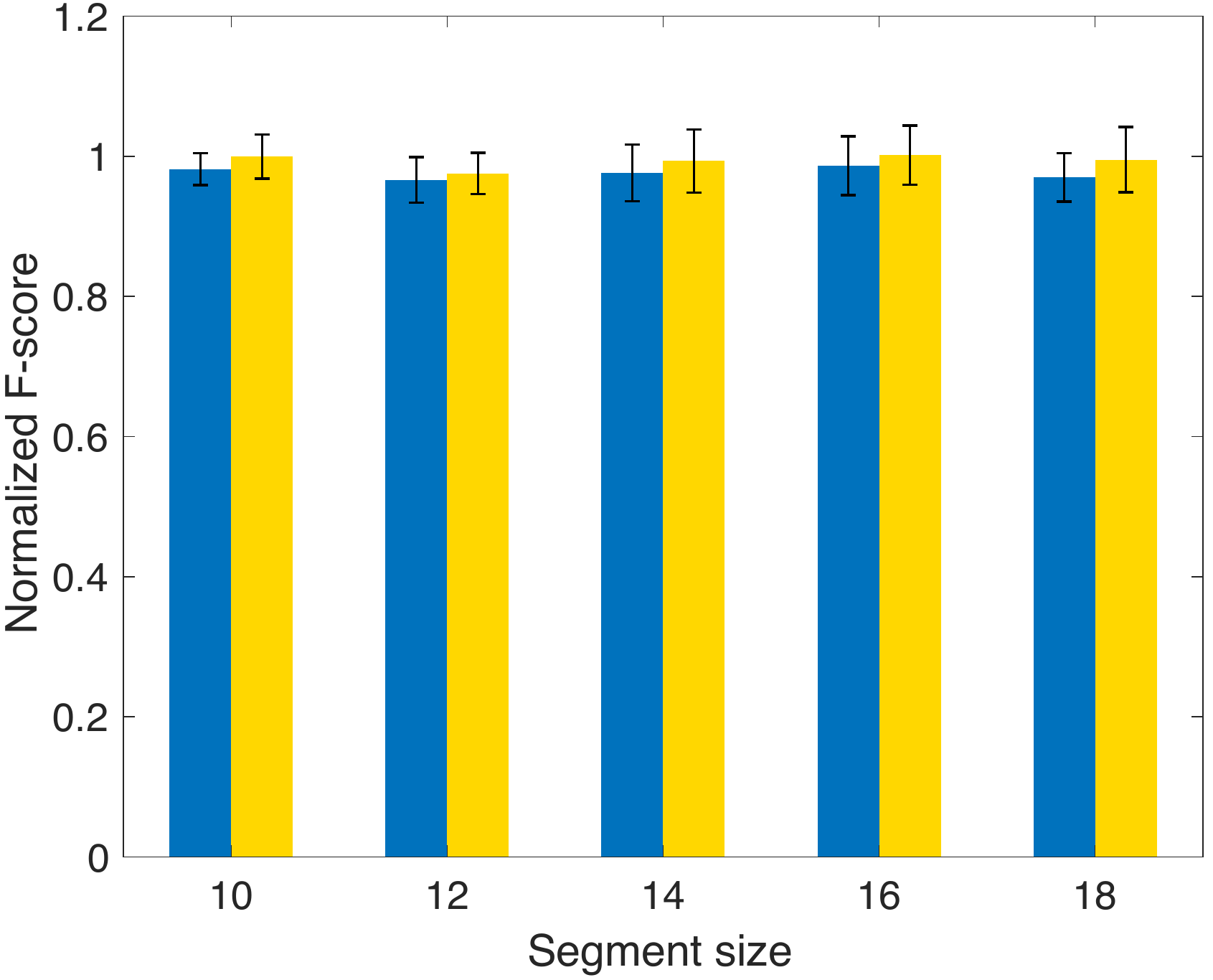}}\hspace{3mm}
	\subfloat[OVP Linear\label{subfig:ovp-speed}]{
		\includegraphics[width=.283\textwidth]{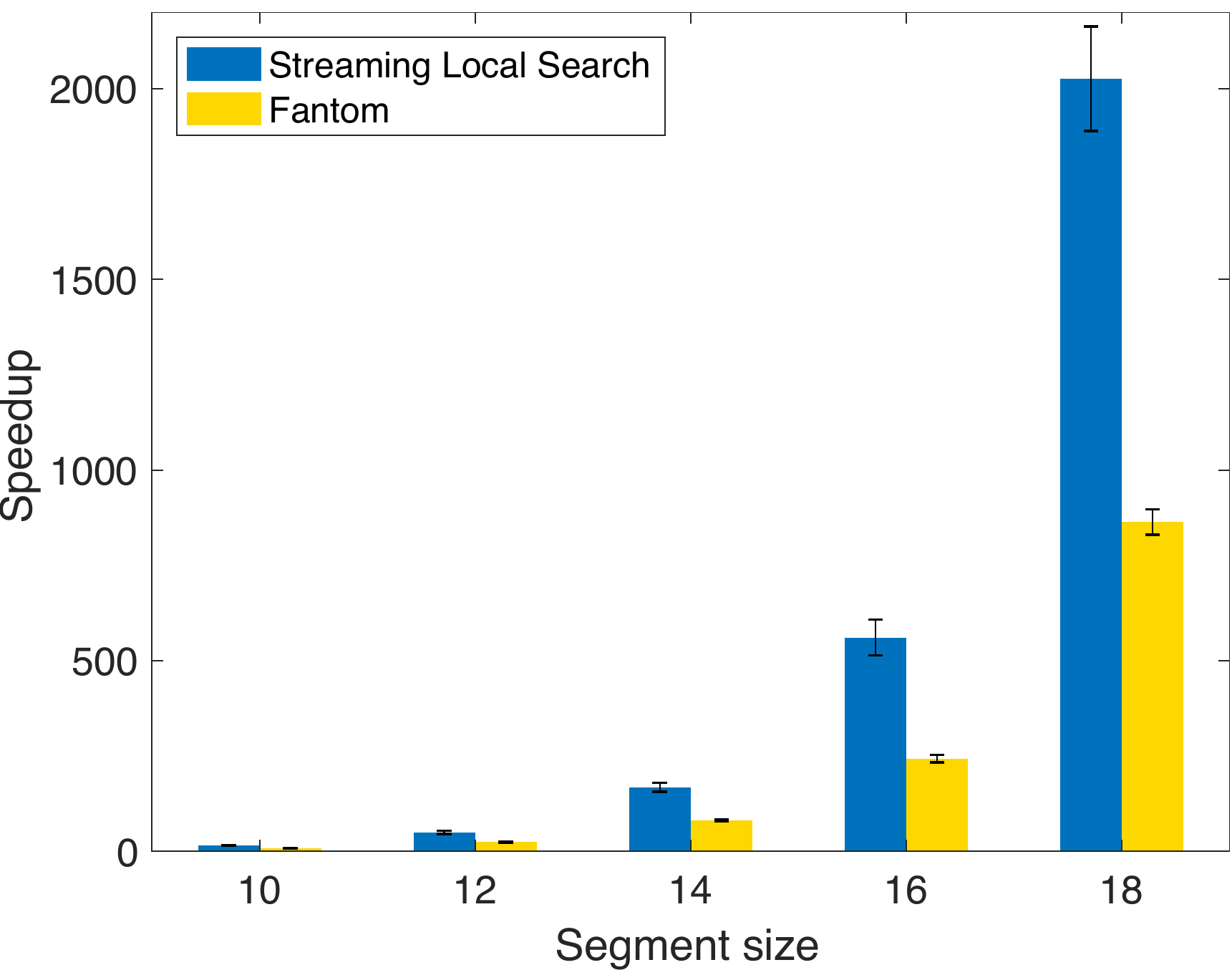}}\hspace{5mm}
	\subfloat[OVP Linear\label{subfig:ovp-linear}]{
		\includegraphics[width=.267\textwidth]{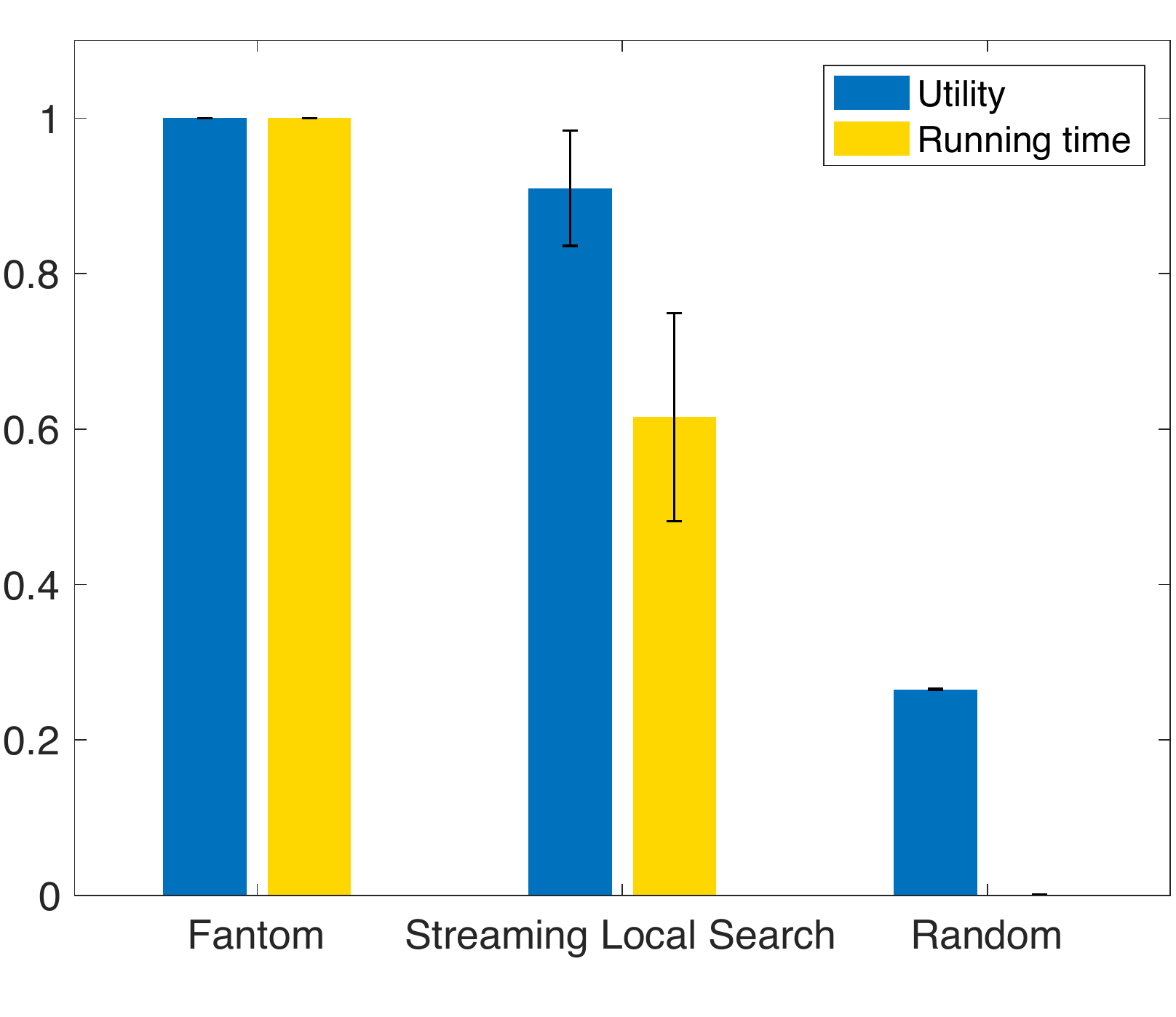}}
	\\
	\subfloat[OVP N. Nets\label{subfig:ovp-nn-fs}]{
		\includegraphics[width=.278\textwidth]{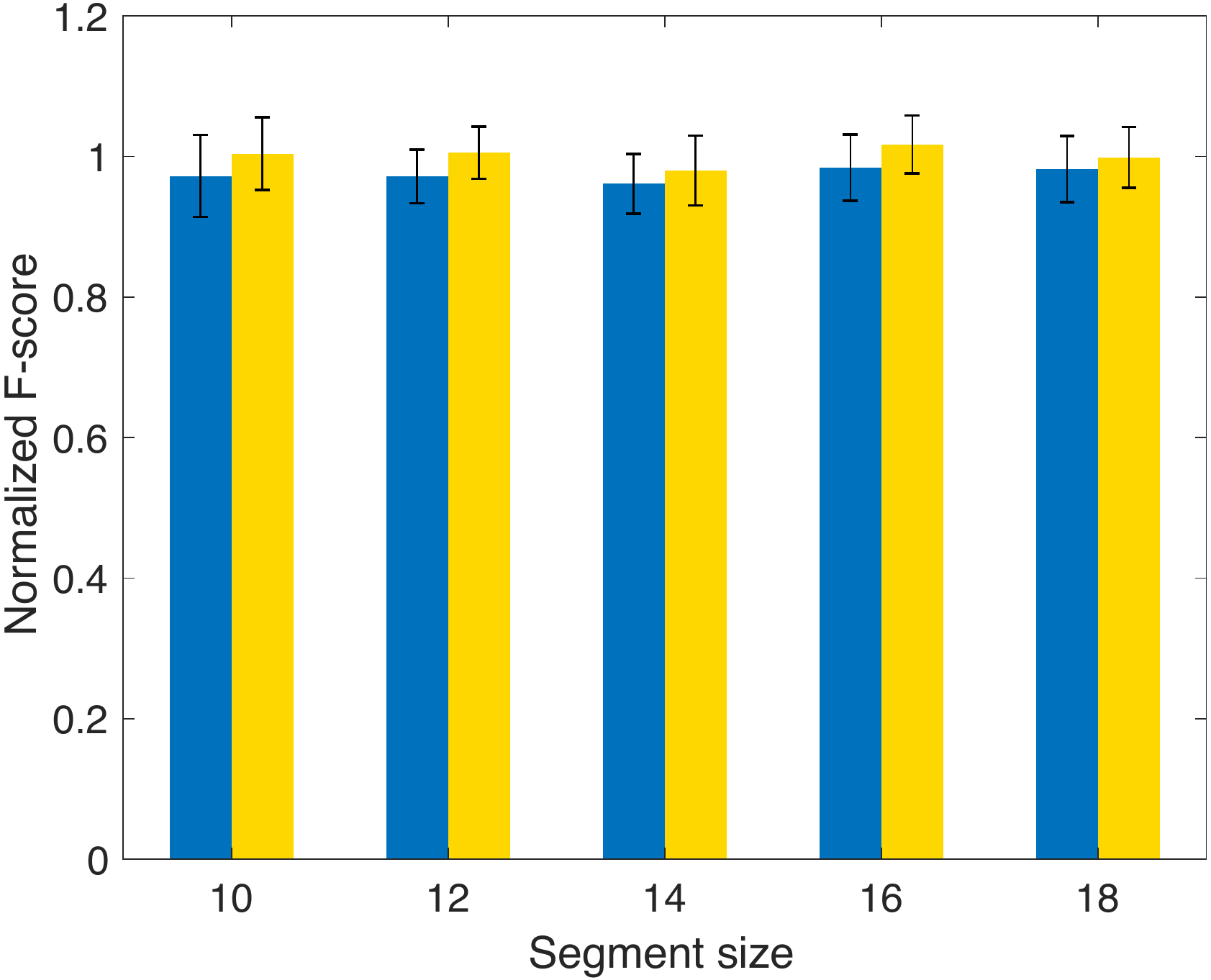}}\hspace{3mm}
	\subfloat[OVP N. Nets\label{subfig:ovp-nn-speed}]{
		\includegraphics[width=.283\textwidth]{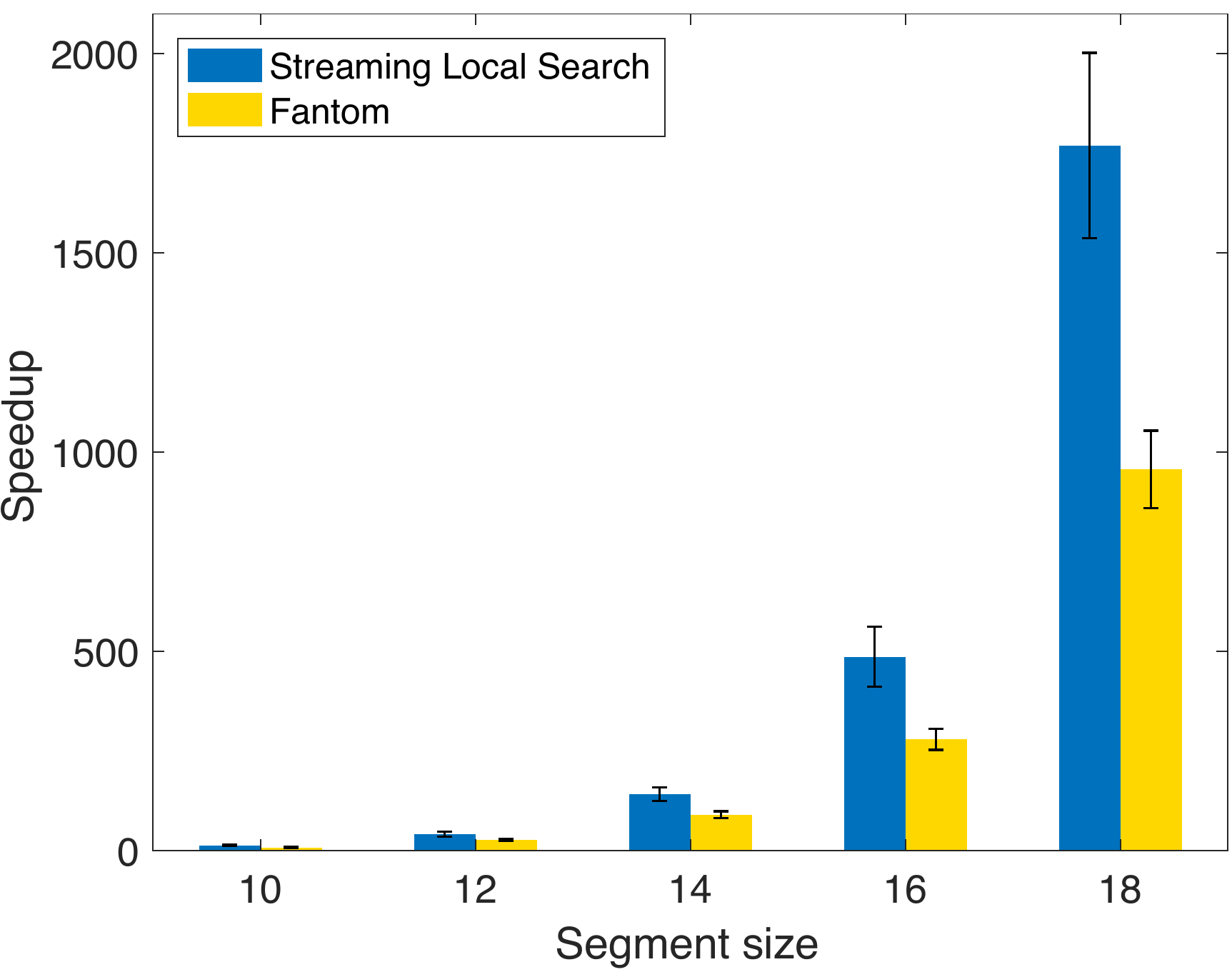}}\hspace{5mm}
	\subfloat[OVP N. Nets\label{subfig:ovp-nn}]{
		\includegraphics[width=.267\textwidth]{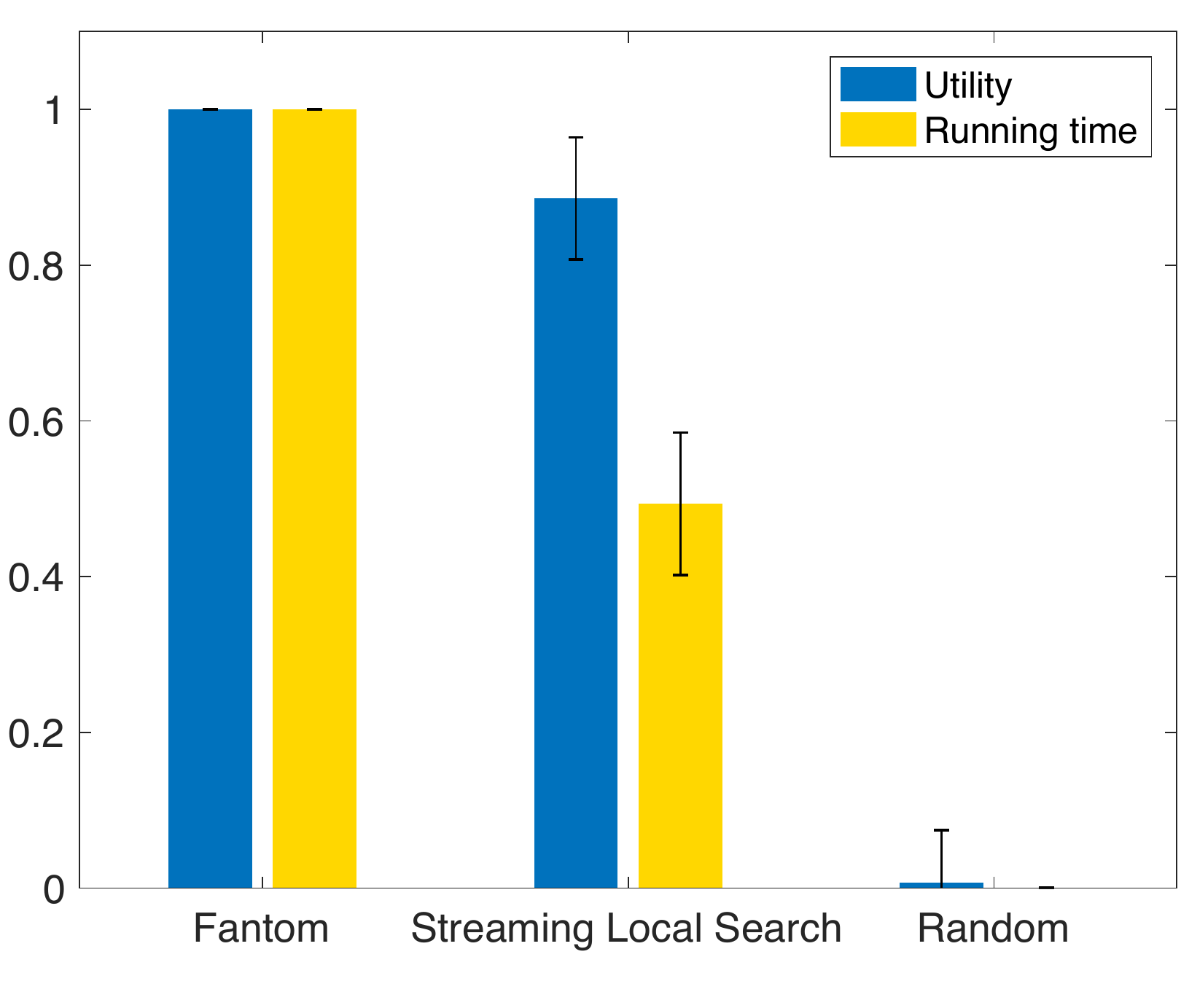}}
	\caption{
			Performance of \SLS compared to the other benchmarks. a), d) show the ratio of the F-score obtained by \SLS and \AlgF vs. the F-score obtained by the method of \protect\citet{gong2014diverse}, using the sequential DPP objective and linear embeddings on YouTube and OVP datasets. g), j) show the relative F-scores for non-linear features from a one-hidden-layer neural network. b), e), h), k) show the speedup of \SLS and \AlgF over the method of \protect\citet{gong2014diverse}. c), f), i), l) show the utility and running time for \SLS and random selection vs. the utility and running time of \AlgF, using the original DPP objective.
	}
\end{figure*}\label{fig:fs-speed}

\subsubsection{Dataset.} For our experiments, we use the Open Video Project (OVP), and the YouTube datasets with 50 and 39 videos, respectively \cite{de2011vsumm}. 
We use the pruned video frames as described in \cite{gong2014diverse}, where one frame is uniformly sampled per second, and uninformative frames are removed.
Each video frame is then associated with a feature vector that consists of Fisher vectors \cite{perronnin2007fisher} computed from SIFT features %$(\phi_i)$ 
\cite{lowe2004distinctive}, contextual features, and features computed from the frame saliency map 
%$(x_i)$
\cite{rahtu2010segmenting}. The size of the feature vectors, $v_i$, are 861 and 1581 for the OVP and YouTube datasets. %, respectively. 

\begin{figure*}[!htb]
	\includegraphics[width=1\textwidth]{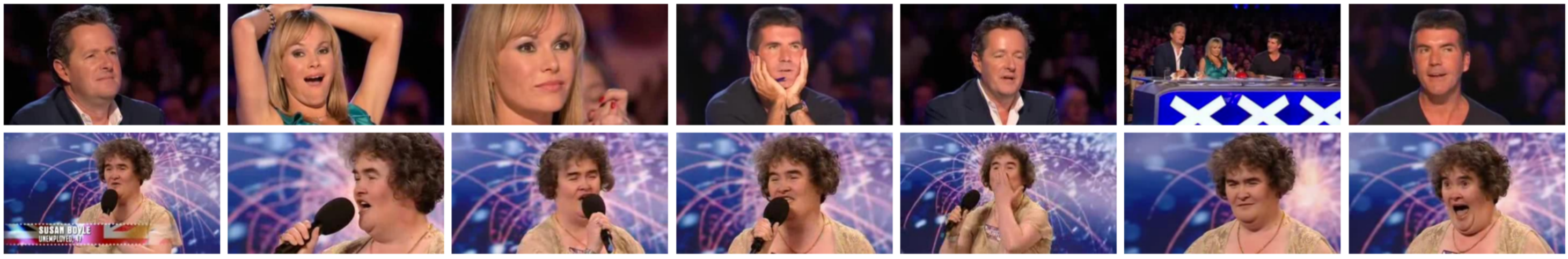}
	\caption{
		Summary produced by \SLS, focused on judges and singer for YouTube video 106.\\
	}\label{fig:v106}
\vspace{-1mm}
\end{figure*}
\begin{figure*}[!thb]
	\includegraphics[width=1\textwidth]{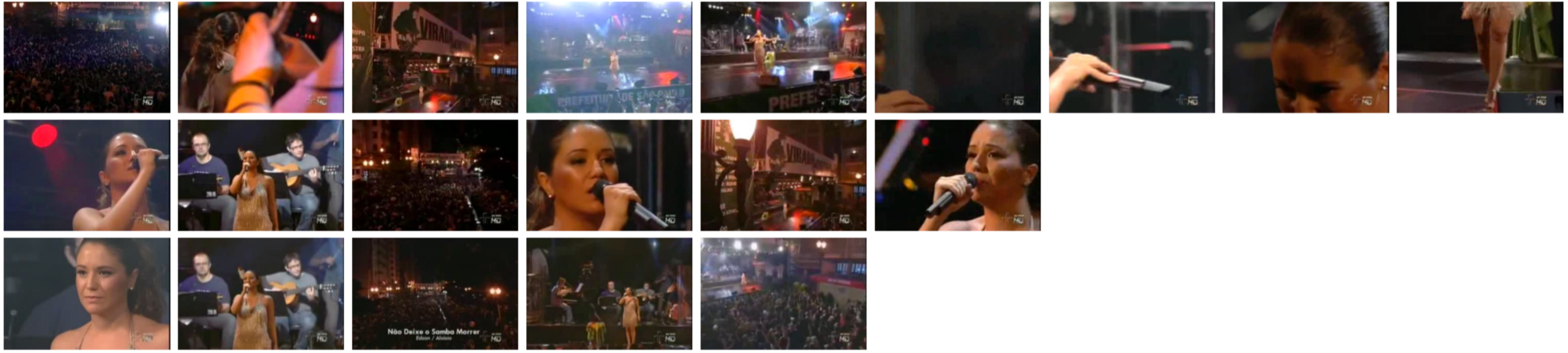}
	\caption{
			Summary produced by method of \protect\citet{gong2014diverse} (top row), vs. \SLS (middle row), and a user selected summary (bottom row), for YouTube video 105.
	}\label{fig:v105}
\vspace{-3mm}
\end{figure*}

The DPP kernel $L$ (Eq.~\ref{eq:dpp}), can be parametrized and learned via maximum likelihood estimation \cite{gong2014diverse}. 
For parametrization, we follow \cite{gong2014diverse}, and
use both a linear transformation,  i.e. $L_{ij}=v_i^T W^T W v_j$, as well as a non-linear transformation using a one-hidden-layer neural network, 
i.e. $L_{ij} = z^T_i W^TW z_j$ where $z_i = \tanh(U v_i)$, and $\tanh(.)$ stands for the hyperbolic transfer function. The parameters, $U$ and $W$ or just $W$, are learned on 80\% of the videos, selected uniformly at random. 
By the construction of \cite{gong2014diverse}, we have $\det(L) > 0$. However, $\det(L)$ can take values less than 1, and the function is non-monotone. 
We added a positive constant to the function values to make them non-negative. 
Following \citet{gong2014diverse} for evaluation, we treat each of the 5 human-created summaries per video as ground truth for each video. 

\subsubsection{Sequential DPP.}
To capture the sequential structure in video data, \citet{gong2014diverse} proposed a sequential DPP. Here, a long video sequence is partitioned into $T$ disjoint yet
consecutive short segments,  
and for selecting a subset $S_t$ from each segment $t\in[1,T]$, a DPP is imposed over the union of the frames in the segment $t$ and the selected subset $S_{t-1}$ in the immediate past frame $t-1$.
The conditional distribution of the selected subset from segment $t$ is thus given by
$
\mathcal{P}(S_t|S_{t-1}) = \frac{\det (L_{S_t \cup S_{t-1}})}{\det(I_t + L_{S_{t-1} \cup V_t})},
$
where $V_t$ denotes all the video frames in segment $t$, and $I_t$ is a diagonal matrix in which the elements corresponding to $S_{t-1}$ are zeros and the elements corresponding to $S_t$ are 1. 
MAP inference for the sequential DPP is as hard as for the standard DPP, but submodular optimization techniques can be used to find approximate solutions. In our experiments, we use a sequential DPP as the utility function in all the algorithms.

\subsubsection{Results.}
Table \ref{table:scores} shows the F-score, Precision and Recall for 
%the three algorithms, 
our algorithm, that of \citet{gong2014diverse} and \AlgF \cite{mirzasoleiman2016fast},
for segment size $|V_t|=10$. It can be seen that in all three metrics, the summaries generated by \SLS are competitive to the two centralized baselines. 

Fig. \ref{subfig:you-fs}, \ref{subfig:ovp-fs} show the ratio of the F-score obtained by \SLS and \AlgF vs. the F-score obtained by exhaustive search \cite{gong2014diverse} for varying segment sizes, using linear embeddings on the YouTube and OVP datasets. 
It can be observed that our streaming method achieves the same solution quality as the centralized baselines. 
Fig. \ref{subfig:you-speed}, \ref{subfig:ovp-speed} show the speedup of \SLS and \AlgF over the method of \citet{gong2014diverse}, for varying segment sizes.
We note that both \AlgF and \SLS obtain a speedup that is exponential in the segment size.
In summary, \SLS achieves solution qualities comparable to \cite{gong2014diverse}, but 1700 times faster than \cite{gong2014diverse}, and 2 times
faster than \AlgF for larger segment size. This makes our streaming method an appealing solution for extracting real-time summaries.
In real-world scenarios, video frames are typically generated at such a fast pace that larger segments make sense. Moreover, unlike the centralized baselines that need to first buffer an entire segment, and then produce summaries, our method generates real-time summaries after receiving each video frame. This capability is crucial in privacy-sensitive applications. 

Fig. \ref{subfig:you-nn-fs} and \ref{subfig:ovp-nn-fs} show similar results for nonlinear representations, where a one-hidden-layer neural network is used to infer a hidden representation for each frame. 
We make two observations: First, non-linear representations generally improve the solution quality. Second, as before, our streaming algorithm achieves exponential speedup (Fig. \ref{subfig:you-nn-speed}, \ref{subfig:ovp-nn-speed}).

Finally, we also compared the three algorithms with a “standard”, non-sequential DPP as the utility function, for generating summaries of length 5\% of the video length. Again, our method yields competitive performance with a much shorter running time (Fig. \ref{subfig:you-linear}, \ref{subfig:you-nn}, \ref{subfig:ovp-linear}, \ref{subfig:ovp-nn}).

\subsubsection{Using constraints to generate customized summaries.}

In our second experiment, we show how constraints can be applied to generate customized summaries.
We apply \SLS to YouTube video 106, which is a part of America's Got Talent series. It features a singer and three judges in the judging panel. Here, we generated two sets of summaries using different constraints. The top row in Fig. \ref{fig:v106} shows a summary focused  on the judges. Here we considered 3 uniform matroid constraints to limit the number of frames chosen containing each of the judges,
i.e., $\cI\!=\!\{S\!\subseteq\! V\!:|S \cap V_j|\leq l_j\}$, where $V_j \!\subseteq \!V$ is the subset of frames containing judge $j$, and $j \in [1,3]$; the $V_j$ can overlap.
The limits for all the matroid constraints are $l_j=3$. To produce real-time summaries while receiving the video, we used the Viola-Jones algorithm \cite{viola2004robust} to detect faces in each frame, and trained a multiclass support vector machine using histograms of oriented gradients (HOG) to recognize different faces. The bottom row in Fig. \ref{fig:v106} shows a summary focused on the singer using one matroid constraint.

To further enhance the quality of the summaries, we assigned different weights to the frames based on the probability for each frame to contain objects, using selective search \cite{uijlings2013selective}. By assigning higher cost to the frames that have low probability of containing objects, 
and by limiting the total cost of the selected elements by a knapsack, we can filter uninformative and blurry frames, and produce a summary closer to human-created summaries. Fig. \ref{fig:v105} compares the result of
our method, the method of \citet{gong2014diverse} and a human-created summary. 

\vspace{-2mm}
\section{Conclusion}
We have developed the first streaming algorithm, \SLS, for maximizing non-monotone submodular functions subject to a collection of independence systems and multiple knapsack constraints. 
In fact, our work provides a general framework for converting monotone streaming algorithms to non-monotone streaming algorithms for general constrained submodular maximization.
We demonstrated its applicability to streaming video summarization with various personalization constraints. Our experimental results show that our method can speed up the summarization task more than 1700 times, while achieving a similar performance as centralized baselines. This makes it a promising approach for many real-time summarization tasks in machine learning and data mining. 
Indeed, our method applies to any summarization task with a non-monotone (nonnegative) submodular utility function, and a collection of independence systems and multiple knapsack constraints. 

\subsubsection{Acknowledgments.} This research was partially supported by  ERC StG 307036, and NSF CAREER 1553284.

\vspace{-1mm}
\fontsize{9.0pt}{10.0pt} \selectfont
\begin{small}
	\bibliography{mirzasoleiman-2198}
	\bibliographystyle{aaai.bst}
\end{small}

%\newpage
%\appendix
% !TEX root = aaai18.tex
\onecolumn
\section*{\centering\Large{Supplementary Materials. }}
\section{Analysis of \SLS}
\paragraph{Proof of theorem~\ref{thm:psys}}
\begin{proof}
Consider a chain of $r$ instances of our streaming algorithm, i.e. $\{\pstream_1, \cdots, \pstream_r \}$. 
For each $i \in [1, r]$, $\pstream_i$ provides an $\alpha$-approximation guarantee on the ground set  $V_i$ of items it has received. Therefore we have:
\begin{align}
f(S_i)\geq \alpha {f(S_i\cup C_i)},%-\frac{\rho d}{4p} . 
\label{eq:si-bound}
\end{align}
%where $\alpha$ is the approximation guarantee for \pstream,
where $C_i = C^* \cap V_i$ for all $i \in [1, r]$, and $C^*$ is the optimal solution. %, and $X_i$ is the subset of elements processed by $\pstream_i$. 
%and hence $C_1 = C^*$. 
Moreover, for each $i$, $S'_i$ is the solution of the unconstrained maximization algorithm on ground set $S_i$. Therefore, we have:
\begin{align}
f(S_i')\geq \beta f(S_i\cap C_i), \label{eq:sip-bound}
\end{align}
where $\beta$ is the approximation guarantee of the unconstrained submodular maximization algorithm (\textsc{Unconstrained-Max}).\\

\noindent We now use the following lemma from \cite{buchbinder2014submodular} to bound the total value of the solutions provided by the $r$ instances of $\pstream$.
%\vspace{-4mm}
\begin{lemma}[\textbf{\textit{Lemma 2.2. of} \cite{buchbinder2014submodular}}]\label{lemma:buch}
	 Let $f' : 2^V \rightarrow R$ be submodular. Denote by $A(p)$ a random subset of A where each element
	appears with probability at most $p$ (not necessarily independently). Then, $\mathbb{E}[f'(A(p))] \geq (1-p)f'(\emptyset)$.
\end{lemma}
%\vspace{-4mm}
\noindent
Let $S$ be a random set which is equal to every one of the sets $\{S_1,\cdots, S_r\}$ with probability $p=1/r$. %Since these sets are disjoint, every element of $V$ belongs to $S$ with probability at most $p =1/k$. 
For $f':2^V \rightarrow R$, and $f'(S)=f(S\cup \opt)$, from Lemma \ref{lemma:buch} we get:
\begin{align} \label{eq:4}
\mathbb{E}[f'(S)]&=\mathbb{E}[f(S \cup C^*)] = \frac{1}{r}\sum_{i=1}^r f(S_i \cup C^*) \stackrel{\text{Lemma \ref{lemma:buch}}}{\geq} (1-p)f'(\emptyset)=(1-\frac{1}{r})f(C^*) 
\end{align}
Also, note that each instance $i$ of \pstream in the chain has processed all the elements of the ground set $V$ except those that are in the solution of the previous instances of \pstream in the chain. As a result, $V_i = V \setminus \cup_{j=1}^{i-1} S_i$, and for every $i \in [1,r]$, we can write:
\begin{align}\label{eq:8}
f(C_i)+ f(C^* \cap (\cup_{j=1}^{i-1}S_j)) = f(C_i)+f(\cup_{j=1}^{i-1} (C^* \cap S_j))=f(C^*).
\end{align}

\noindent Now, using Eq. \ref{eq:4}, and via a similar argument as used in \cite{feldman2017greed}, we can write:
\begin{align}
(r-1)f(C^*) & \leq \sum_{i=1}^r f(S_i \cup C^*) &\textrm{By Eq. \ref{eq:4}} \nonumber\\
%& \leq  \sum_{i=1}^t f(S_i \cup C_i) + \sum_{i=1}^t f(C^* \setminus X_i)\\ \nonumber
& \leq \sum_{i=1}^r \left[ f(S_i \cup C_i) + f\big(\cup_{j=1}^{i-1} (C^* \cap S_j)\big) \right]\label{eq:3}&\textrm{By Eq. \ref{eq:8}}\\ 
& \leq \sum_{i=1}^r \bigg[ f(S_i \cup C_i) +  \sum_{j=1}^{i-1} f(C^* \cap S_j) \bigg] \label{eq:6} \\ 
& \leq \sum_{i=1}^r \bigg[\frac{1}{\alpha} f(S_i) + \frac{1}{\beta} \sum_{j=1}^{i-1} f(S'_j) \bigg] &\textrm{By Eq. \ref{eq:si-bound}, Eq. \ref{eq:sip-bound}}\nonumber\\
& \leq  \sum_{i=1}^r \bigg[ \frac{1}{\alpha} f(S) + \frac{1}{\beta} \sum_{j=1}^{i-1} f(S) \bigg]&\textrm{By definition of $S$ in Algorithm \ref{alg:psys}}\nonumber\\
& = \left(\frac{r}{\alpha} + \frac{r(r-1)}{2\beta}\right)f(S).\nonumber
\end{align}
%where Eq. \ref{eq:3} follows from the fact that for every $i \in [1,r]$, we have:
%$$f(C_i)+ f(C^* \cap (\cup_{j=1}^{i-1}S_i)) = f(C_i)+f(\cup_{j=1}^{i-1} (C^* \cap S_i))=f(C^*).$$
\noindent Hence, we get:
\begin{equation}\label{eq:aprx}
f(S) \geq \frac{r-1}{r/\alpha+ r(r-1)/2\beta}f(C^*)
\end{equation}

\noindent Taking the derivative w.r.t. $r$, we get that the ratio is maximized for $r=\left\lceil\sqrt{\frac{2\beta}{\alpha}}+1\right\rceil$. Plugging this value into Eq. \ref{eq:aprx},  we have:
\begin{align*}
f(S) & \geq
\frac{1-\frac{1}{\sqrt{2\beta/\alpha}+1}}{\frac{1}{\alpha}+\frac{\sqrt{2\beta/\alpha}}{2\beta}} f(C^*)\\\nonumber
& = \frac{\sqrt{2\beta/\alpha}}{(\sqrt{\frac{2\beta}{\alpha}}+1)(\frac{1}{\alpha} + \frac{\sqrt{2\beta/\alpha}}{2\beta})}f(C^*)\\\nonumber
& = \frac{\sqrt{2\beta}}{(\sqrt{2\beta}+1/\sqrt{\alpha})(1/\sqrt{\alpha} + 1/\sqrt{2\beta})}f(C^*)\\\nonumber
& = \frac{\sqrt{2\beta}}{(1/\sqrt{\alpha}+1/\sqrt{2 \beta})^2}f(C^*)\\\nonumber
\end{align*}

\noindent Using $\beta=1/2$ from \cite{buchbinder2015tight}, we get the desired result:
$$f(S) \geq \frac{1}{(1/\sqrt{\alpha}+1)^2}f(C^*)$$

\noindent Finally, Corollary \ref{col:psys} follows by replacing $\alpha=1/4p$ from \cite{chekuri2015streaming} and $\beta = 1/2$ from \cite{buchbinder2015tight}:
\begin{equation*}
f(S) \geq \frac{1}{(2\sqrt{p}+1)^2}f(C^*)
\end{equation*}
\end{proof}

\noindent For calculating the average update time, we consider the worst case scenario, where every element can go through the entire chain of $r$ instances of \pstream at some point during the run of \SLS. Here the total running time of the algorithm is $O(nrT)$, where $n$ is the size of the stream, and $T$ is the update time of \pstream. Hence the average update time per element for \SLS is $O(nrT/n)=O(rT)$.

\paragraph{Proof of theorem~\ref{thm:pknapsack}}
\begin{proof}
Here, a (fixed) density threshold $\rho$ is used to restrict the \pstream to only pick elements if $\frac{f_{S_i}(e)}{\sum_{j=1}^d c_{je}}\geq \rho$. We first bound the approximation guarantee of this new algorithm \pdstream, and then use a similar argument as in the proof ot Theorem \ref{thm:psys} to provide the guarantee for \SLS.
	Consider an optimal solution $C^*$ and set: 
	\begin{equation}\label{eq:rhoo}	
	\rho^* = \frac{2}{\left(\frac{1}{\sqrt{\alpha}}+\frac{1}{\sqrt{\beta}}\right)\left(\frac{1}{\sqrt{\alpha}}+2d\sqrt{\alpha}+\frac{1}{\sqrt{\beta}}\right)} f(C^*).
	\end{equation}
	 %$\rho^* ={2\cdot (4p-1)\cdot f(C^*)/4p(8p+2d-1)}$. 
	By submodularity we know that $m \leq f(C^*) \leq m k$,  where $k$ is an upper bound on the cardinality of the largest feasible solution, and $m$ is the maximum value of any singleton element. Hence:
	\begin{align*}
	\frac{2m}{\left(\frac{1}{\sqrt{\alpha}}+\frac{1}{\sqrt{\beta}}\right)\left(\frac{1}{\sqrt{\alpha}}+2d\sqrt{\alpha}+\frac{1}{\sqrt{\beta}}\right)} \leq \rho^* \leq \frac{2mk}{\left(\frac{1}{\sqrt{\alpha}}+\frac{1}{\sqrt{\beta}}\right)\left(\frac{1}{\sqrt{\alpha}}+2d\sqrt{\alpha}+\frac{1}{\sqrt{\beta}}\right)}.
	\end{align*}
	Thus there is a run of the algorithm with density threshold $\rho \in R$ such that: 
	\begin{equation}\label{eq:density}
	\rho\leq \rho^*\leq (1+\epsilon)\rho.
	\end{equation}
	 
	%For this run of the algorithm, we have the following.
	%%%
	%\sj{I don't understand. What is $k$ vs $r$? Also, as typed and with these definitions, the above inequality does not hold. Explain what you are trying to do, and double check.}
	\noindent
	For the run of the algorithm corresponding to $\rho$, we call the solution of the first instance $\pdstream_1$, $S_\rho$.
	If $\pdstream_1$ terminates by exceeding some knapsack capacity, we know that for one of the knapsacks $j \in [d]$, we have  $c_j(S_\rho)>1$, and hence also $\sum_{j = 1}^d c_j(S_\rho) > 1$ (W.l.o.g. we assumed the knapsack capacities are 1).
	On the other hand, the extra density threshold we used for selecting the elements tells us that for any $e \in S_\rho$, we have $\frac{f_{S_\rho}(e)}{\sum_{j=1}^d c_{je}} \geq \rho$. 
	I.e., the marginal gain of every element added to the solution $S_\rho$ was greater than or equal to $\rho \sum_{j=1}^d c_{je} $.
	Therefore, we get:
	\begin{align*}
	%f(S_\rho) = \sum_{j=1}^{|S_\rho|} f_{S_\rho}(j) \geq {\rho} \sum_{j \in S_\rho} \sum_{i \in [d]} c_{i,j} > {\rho}.
	f(S_\rho) \geq  \sum_{e \in S_\rho}  \big(\rho \sum_{j =1}^d c_{je} \big) > {\rho}.
	\end{align*}
	%\sj{What exactly is $S$ here? This is in fact also a bit confusing in the algorithm. The set of the previous iteration?}
	Note that $S_\rho$  is not a feasible solution, as it exceeds the $j$-th knapsack capacity.
	However, the solution before adding the last element $e$ to $S_\rho$, i.e. $T_\rho = S_\rho - \{e\}$, and the last element itself are both feasible solutions, and by submodularity, the best of them provide us with the value of at least $$\max\{f(T_\rho), f(\{e_f\})\} \geq \frac{\rho}{2}. $$%\geq \frac{2(1-\alpha)}{(1+\eps)(2/\alpha+2d-1)}f(C^*)$$
	
	\noindent On the other hand, if $\pdstream_1$ terminates without exceeding any knapsack capacity, 
	%as it cannot add items due to $p$-system constraint and never exceeds any knapsack capacity. Let $S_\rho$ be the set returned, and $C \in \mathcal{I}$ be any feasible solution to the problem. 
	we divide the elements in $C^* \setminus S_\rho$ into two sets.
	Let $C^*_{<\rho}$ be the set of elements from $C^*$ which cannot be added to $S_\rho$ because their density is below the threshold, i.e., $\frac{f_{S_\rho}(e)}{\sum_{i=1}^d c_{je}}< \rho$ and $C^*_{\geq \rho}$ be the set of elements from $C^*$ which cannot be added to $S_\rho$ due to independence system constraints.
	 %Then we first derive inequalities with respect to each of these sets. 
	%First consider the set $C_{<\rho}$. %We have two cases:
	For the elements of the optimal solution $C^*$ which cannot be added to $S_\rho$ because their density is below the threshold, we have:
	
	%\vspace{-7.5mm}
	\begin{small}
		\begin{eqnarray*}
		f_{S_\rho}(C^*_{<\rho})%&\leq& \sum_{e\in C_{<\rho}}f_{S_{\rho}}(e) %\nonumber \\[-3mm]   \textrm{ (By submodularity)}
		\leq \sum_{e\in C_{<\rho}} {\rho} \sum_{j=1}^d c_{je} %\nonumber \\[-3mm]  \textrm{ (By definition }C_{<\rho}) 
		= \rho \sum_{j=1}^d \sum_{e\in C_{<\rho}} c_{je} %\nonumber \\[-3mm]
		%\leq \rho \sum_{i=1}^d 1  %\nonumber \\[-1mm] \textrm{  is a feasible solution)}  \textrm{ (As }C_{<\rho}
		%\leq {d\rho}
		%\leq d\rho^* %= \frac{2 d(1-\alpha)}{2/\alpha+2d-1}f(C^*)
		%
		\end{eqnarray*}
	\end{small}
	
	\noindent Since $C_{<\rho}$ is a feasible solution, we know that $\sum_{e\in C_{<\rho}} c_{je} \leq 1$, and therefore:
	\begin{equation}\label{eq:greater-than-rho}
	f_{S_\rho}(C^*_{<\rho}) \leq {d\rho}
	\leq \rho \sum_{j=1}^d \sum_{e\in C_{<\rho}} c_{je}
	\leq d\rho
	\leq d\rho^*
	\end{equation}
	%\vspace{-9mm}
	On the other hand, 
	if the ground set was restricted to elements that pass the density threshold, then $S_\rho$ would be a subset of that ground set, and the approximation guarantee of $\pstream_1$ still holds; hence
	%since $S_\rho$ is a feasible solution for $\pstream_1$ without any density threshold, 
	from Eq. \ref{eq:si-bound} we know that: $$f(S_\rho) \geq  \alpha f(S_\rho \cup C^*_{\geq \rho}),$$ and thus we obtain:
	%\vspace{-2mm}
	\begin{align}
	f_{S_\rho}(C^*_{\geq \rho})=& f(S_\rho \cup C^*_{\geq \rho})-f(S_\rho) \leq \big(\frac{1}{\alpha}-1\big) f(S_\rho). %= \frac{1-\alpha}{\alpha} f(S_\rho).
	 \label{eq:less-than-rho}
	\end{align}
	Adding Eq~\ref{eq:greater-than-rho} and \ref{eq:less-than-rho}, and using submodularity we get:
	\begin{align*}
	%f_{S_\rho}(C^*) \leq& f_{S_\rho}(C^*_{<\rho})\!+\!f_{S_{\rho}}(C^*_{\geq \rho})
	f(S_\rho \cup C^*)-f(S_\rho) \leq& f_{S_\rho}(C^*_{<\rho})\!+\!f_{S_{\rho}}(C^*_{\geq \rho})
	\leq  \big(\frac{1}{\alpha}-1\big) f(S_\rho) +  d \rho % \frac{2\alpha(1-\alpha)}{2-\alpha+2d}f(C^*). 
	%\Rightarrow f(S_\rho)\geq& \frac{1}{p+1}f(S_\rho\cup C)-\frac{\rho \ell}{p+1}
	\end{align*}
	Therefore,
	\begin{align}\label{eq:knapsack}
	f(S_\rho) \geq \alpha f(S_\rho \cup C^*)-\alpha d \rho.
	\end{align}
Now, using a similar argument as in the proof of Theorem \ref{thm:psys}, we have:
\begin{align*}
(r-1)f(C^*) & \leq \sum_{i=1}^r f(S_i \cup C^*) &\textrm{By Eq. \ref{eq:4}}\\
	& \leq \sum_{i=1}^r f(S_i \cup C_i) + \sum_{i=1}^r \sum_{j=1}^{i-1} f(C^* \cap S_j)&\textrm{By Eq. \ref{eq:6}}\\\nonumber
	& \leq \frac{1}{\alpha} \sum_{i=1}^r [f(S_i) + \alpha d\rho] + \frac{1}{\beta} \sum_{i=1}^r \sum_{j=1}^{i-1} f(S'_j) &\textrm{By Eq. \ref{eq:knapsack}}\\\nonumber	
	& \leq \frac{1}{\alpha} \sum_{i=1}^r [f(S) + \alpha d\rho] + \frac{1}{\beta} \sum_{i=1}^r \sum_{j=1}^{i-1} f(S) &\textrm{By definition of $S$ in Algorithm \ref{alg:pknapsack}}\\\nonumber
	& = \left(\frac{r}{\alpha} + \frac{r(r-1)}{2\beta}\right)f(S)+rd\rho
\end{align*}
\end{proof}
\noindent
Hence, we have:
\begin{equation*}\label{eq:apx}
f(S) \geq \frac{r-1}{r/\alpha+ r(r-1)/2\beta}f(C^*) - \frac{rd\rho}{r/\alpha+ r(r-1)/2\beta}f(C^*)
\end{equation*}
From Eq. \ref{eq:density}, we know that $\rho \geq (1-\eps)\rho^*$. Using Eq. \ref{eq:rhoo}, we get:
\begin{align*}
f(S) & \geq \frac{r-1}{r/\alpha+ r(r-1)/2\beta}f(C^*) - \frac{
	\frac{2rd (1-\eps)}{
	(1/\sqrt{\alpha}+1/\sqrt{\beta})(1/\sqrt{\alpha}+2d\sqrt{\alpha}+1/\sqrt{\beta})}
}
{r/\alpha+ r(r-1)/2\beta}f(C^*)\\\nonumber
%& =
%\frac{1-\frac{1}{\sqrt{2\beta/\alpha}+1} 
%	- 2 d/\big[ (1+\frac{1}{\sqrt{\alpha}})(1+\frac{2d+1}{\sqrt{\alpha}})\big]}{\frac{1}{\alpha}+\frac{\sqrt{2\beta/\alpha}}{2\beta}}f(C^*)\\\nonumber
%& = \frac{
%	\big[1-\frac{1}{\sqrt{2\beta/\alpha}+1}\big]
%	\big[ (1+\frac{1}{\sqrt{\alpha}})(1+\frac{2d+1}{\sqrt{\alpha}})\big] 
%	- 2 d }{
%	\big[\frac{1}{\alpha}+\frac{\sqrt{2\beta/\alpha}}{2\beta}\big]
%	\big[ (1+\frac{1}{\sqrt{\alpha}})(1+\frac{2d+1}{\sqrt{\alpha}})\big]}f(C^*)\\\nonumber
%& = \frac{
%	\big[1-\frac{1}{\sqrt{2\beta/\alpha}+1}\big]
%	\big[ (1+\frac{1}{\sqrt{\alpha}})(1+\frac{2d+1}{\sqrt{\alpha}})\big] 
%	- 2 d }{
%	\big[\frac{1}{\alpha}+\frac{\sqrt{2\beta/\alpha}}{2\beta}\big]
%	\big[ (1+\frac{1}{\sqrt{\alpha}})(1+\frac{2d+1}{\sqrt{\alpha}})\big]}f(C^*)\\\nonumber
\end{align*}

\noindent Plugging in $r=\left \lceil \sqrt{\frac{2\beta}{\alpha}}+1 \right \rceil$ and simplifying, we get  the desired result:
\begin{align*}
f(S) &\geq \frac{\sqrt{\frac{2\beta}{\alpha}}
	- \frac{2d \left(\sqrt{\frac{2\beta}{\alpha}}+1\right)(1-\eps)}
	{ 
	\left(\frac{1}{\sqrt{\alpha}}+\frac{1}{\sqrt{\beta}}\right)\left(\frac{1}{\sqrt{\alpha}}+2d\sqrt{\alpha}+\frac{1}{\sqrt{\beta}}\right)}}
{\frac{1}{\alpha}\sqrt{\frac{2\beta}{\alpha}}
	+\frac{2}{\alpha}
	+\sqrt{\frac{1}{2\beta\alpha}}
	}f(C^*)\\\nonumber
%& = \frac{\sqrt{\frac{2\beta}{\alpha}}
%	[(\frac{1}{\sqrt{\alpha}}+)(\frac{1}{\sqrt{\alpha}}+2d\sqrt{\alpha}+\frac{1}{\sqrt{\beta}}))] - 2d (\sqrt{\frac{2\beta}{\alpha}}+1)}
%	{[\frac{1}{\alpha}\sqrt{\frac{2\beta}{\alpha}}
%	+\frac{2}{\alpha}
%	+\sqrt{\frac{1}{2\beta\alpha}}
%	]
%	[(\frac{1}{\sqrt{\alpha}}+)(\frac{1}{\sqrt{\alpha}}+2d\sqrt{\alpha}+\frac{1}{\sqrt{\beta}}))]}\\\nonumber
& = \frac{\sqrt{2\beta}
	\left(\frac{1}{\sqrt{\alpha}}+\frac{1}{\sqrt{\beta}}\right)\left(\frac{1}{\sqrt{\alpha}}+2d\sqrt{\alpha}+\frac{1}{\sqrt{\beta}}\right) - 2d(1-\eps) \left(\sqrt{2\beta}+\sqrt{\alpha}\right)}
{\left(\frac{\sqrt{2\beta}}{\alpha}
+\frac{2}{\sqrt{\alpha}}
+\sqrt{\frac{1}{2\beta}}
\right)
\left(\frac{1}{\sqrt{\alpha}}+\frac{1}{\sqrt{\beta}}\right)\left(\frac{1}{\sqrt{\alpha}}+2d\sqrt{\alpha}+\frac{1}{\sqrt{\beta}}\right)}f(C^*) \\\nonumber
& \geq \frac{1-\eps}{(1/\sqrt{\alpha}+1/\sqrt{\beta})(1/\sqrt{\alpha}+2d\sqrt{\alpha}+1/\sqrt{\beta})}f(C^*)
\end{align*}

\noindent For $\beta=1/2$ from \cite{buchbinder2015tight}, we get the desired result:
$$f(S) \geq \frac{1-\epsilon}{(1+1/\sqrt{\alpha})(1+2d\sqrt{\alpha}+1/\sqrt{\alpha})}f(C^*)$$

\noindent Corollary \ref{col:knapsack} follows by replacing $\alpha=1/4p$ from \cite{chekuri2015streaming} and $\beta = 1/2$ from \cite{buchbinder2015tight}:
\begin{equation*}
f(S) \geq \frac{1-\eps}{1+4p+4\sqrt{p}+d(2+1/\sqrt{p})}f(C^*)
\end{equation*}

\noindent The average update time for one run of the algorithm corresponding to a $\rho \in R$ can be calculated as in the proof of Theorem \ref{thm:psys}. We run the algorithm for $\log(k)/\eps$ different values of $\rho$, and hence the average update time of \SLS per element is $O(r T \log(k)/\eps)$. %Upon receiving a new element from the stream, is can be processed in parallel for all the $\log(k)/\eps$ values of $\rho$ in parallel. Therefore, using parallel processing, the average update time per element is $O(rT)$.
However, the algorithm can be run in parallel for the $\log(k)/\eps$ values of $\rho$ (line 7 of Algorithm \ref{alg:pknapsack}), and hence using parallel processing, the average update time per element is $O(rT)$.
\end{document}